\documentclass[12pt]{article}
\usepackage{color}
\usepackage{amsmath,amssymb,amsfonts}  
\usepackage{cite}

\newtheorem{definition}{Definition}[section]

\newtheorem{theorem}[definition]{Theorem}

\newtheorem{lemma}[definition]{Lemma}

\newtheorem{rmk}{Remark}[section]

\numberwithin{equation}{section}

\newcommand{\nonu}{\nonumber \\} 

\newcommand{\hs}[1]{\hspace{#1 mm}}

\def\cA{{\cal A}}   \def\cB{{\cal B}}   
\def\cD{{\cal D}}      \def\cF{{\cal F}}  
      \def\cI{{\cal I}}
\def\cJ{{\cal J}}   \def\cK{{\cal K}}   
      
      \def\cR{{\cal R}}
      \def\cU{{\cal U}}
\def\cV{{\cal V}}      
\def\cY{{\cal Y}}     
\def\fD{{\mathfrak D}}

\def\fa{{\mathfrak a}}
\def\fb{{\mathfrak b}}
\def\fc{{\mathfrak c}}
\def\fe{{\mathfrak e}}
\def\ff{{\mathfrak f}}
\def\fg{{\mathfrak g}}
\def\fh{{\mathfrak h}}

\def\fm{{\mathfrak m}}
\def\fn{{\mathfrak n}}

\def\fw{{\mathfrak w}}
\newcommand{\BB}{\mbox{${\mathbb B}$}}
\newcommand{\CC}{{\mathbb C}}
\newcommand{\DD}{{\mathbb D}}

\newcommand{\II}{{\mathbb I}}

\newcommand{\RR}{\mbox{${\mathbb R}$}}

\newcommand{\ZZ}{{\mathbb Z}}

\newcommand{\wh}[1]{\widehat{#1}}
\newcommand{\wt}[1]{\widetilde{#1}}
\newcommand{\wb}[1]{\overline{#1}}
\newcommand{\mb}[1]{\hs{4}\mbox{#1}\hs{4}}

\newcommand{\prf}{\underline{Proof:}\ }
\newcommand{\finprf}{\null \hfill {\rule{5pt}{5pt}}\\ \indent}
\newcommand{\finprfbis}{\null \hfill {\rule{5pt}{5pt}}}
\newcommand{\id}{\mbox{id}}
\newcommand{\atopn}[2]{\genfrac{}{}{0pt}{}{#1}{#2}}
\newcommand{\bs}{\boldsymbol}
\newcommand{\inv}{{\bs{\iota}}}

\definecolor{brique}{rgb}{.9,.2,0}
\definecolor{blvert}{rgb}{0,.8,.85}
\definecolor{vertcl}{rgb}{0,1,.7}
\newcommand\vertcl[1]{\textcolor{vertcl}{#1}}
\newcommand\blvert[1]{\textcolor{blvert}{#1}}
\newcommand\brique[1]{\textcolor{brique}{#1}}
\def\lapth{
\begin{picture}(136,70)(0,-15)\thicklines
\put(0,0){\vertcl{\rule{20pt}{4pt}}}
\put(19,1){\vertcl{\line(1,3){23}}} 
\put(20,1){\vertcl{\line(1,3){23}}} 
\put(21,1){\vertcl{\line(1,3){23}}}
\put(22,1){\vertcl{\line(1,3){23}}}
\put(45,70){\vertcl{\line(1,-3){23}}} 
\put(44,70){\vertcl{\line(1,-3){23}}} 
\put(43,70){\vertcl{\line(1,-3){23}}}
\put(42,70){\vertcl{\line(1,-3){23}}}
\put(2,24){\vertcl{\rule{120pt}{4pt}}}
\put(65,0){\vertcl{\rule{60pt}{4pt}}}
\put(5,37){\Huge{\brique{\textbf{L}}}} 
\put(62,37){\Huge{\brique{\textbf{PTh}}}}
\put(12,-8){\blvert{\rule{92pt}{3.5pt}}}
\put(24,-15){\blvert{\rule{57pt}{3.5pt}}}
\put(36,-22){\blvert{\rule{30pt}{3.5pt}}}
\end{picture}
\raisebox{35pt}{
\begin{minipage}{330pt}\begin{center}
\textbf{Laboratoire d'Annecy-le-Vieux de Physique
Th\'eorique}\\[4ex]
website: \texttt{http://lappweb.in2p3.fr/lapth-2005/}
\end{center}
\end{minipage}}\\
\vspace{10pt}\quad \hrulefill\\
\vspace{10pt}}

\newcommand{\beq}{\begin{equation}}
\newcommand{\eeq}{\end{equation}}
\newcommand{\ben}{\begin{eqnarray}}
\newcommand{\een}{\end{eqnarray}}

\begin{document}
%%%%%%%%%%%%%%%%%%%%%%%%%%%%%%%%
%%%%%  HEADINGS POUR DRAFT  %%%%%%

\pagestyle{empty}
\setcounter{page}{0}
\hspace{-1cm}\lapth

\vfill\vfill
\begin{center}
{\LARGE  {\sffamily 
Nested Bethe ansatz for `all' open spin chains\\[1.2ex]
 with diagonal boundary conditions}}\\[4.2ex]
  
{\large S. Belliard\footnote{samuel.belliard@lapp.in2p3.fr} and E. 
Ragoucy\footnote{eric.ragoucy@lapp.in2p3.fr}
\\[2.1ex]

\textsl{Laboratoire de Physique Th{\'e}orique LAPTH\\[1.2ex]
UMR 5108 du CNRS, associ{\'e}e {\`a} l'Universit{\'e} de
Savoie\\[1.2ex]
 BP 110, F-74941  Annecy-le-Vieux Cedex, France. } }
\end{center}
\vfill\vfill
\begin{abstract}
We present in an unified and detailed way the nested Bethe ansatz for open spin chains 
based on $\cY(gl(\fn))$, $\cY(gl(\fm|\fn))$,
$\wh\cU_{q}(gl(\fn))$ or $\wh\cU_{q}(gl(\fm|\fn))$ (super)algebras, with 
arbitrary representations (i.e. `spins') on each site of the chain 
and diagonal boundary matrices $(K^+(u),K^-(u))$. 
The nested Bethe anstaz applies for a general $K^-(u)$, but a 
particular form of the $K^+(u)$ matrix. 

The construction extends and unifies the results already obtained 
for open spin chains 
based on fundamental representation and for some particular 
super-spin chains. We give the eigenvalues, Bethe equations and 
the explicit form of the Bethe vectors for the corresponding models.
The Bethe vectors are also expressed using a  trace formula.
\end{abstract}

\vfill

\begin{center}
MSC: 81R50, 17B37 ---
PACS: 02.20.Uw, 03.65.Fd, 75.10.Pq
\end{center}

\vfill

\rightline{LAPTH-1309/09}
\rightline{\texttt{arXiv:0902.0321[math-ph]}}
\rightline{February 2009}

\newpage
\markright{\today \dotfill DRAFT\dotfill }
\pagestyle{plain}

\section{Introduction}

The systematic studies of the open spins chains  using  $R$ 
matrices formalism start 
with the seminal papers of Cherednik \cite{Ch} and Sklyanin \cite{S}, who generalized to these 
models the QISM approach developped by the Leningrad school. 
They introduced the 
reflection algebra as the fundamental
ingredient to construct the Abelian Bethe subalgebra and ensure  
integrability of the model.  
This algebra is a subalgebra of the FRT algebra introduced by 
Leningrad group for the periodic spin
chains (for a review, see e.g. \cite{Fad} and references therein). 
The boundary conditions are encoded in two matrices, $K^-(u)$ 
solution of the reflection 
equation, see equation (\ref{ERK}) below,
and $K^+(u)$ solution of the dual equation, see equation (\ref{ERD}).
 With these matrices and the 
standard closed spin chain monodromy matrix, one can construct a transfer 
matrix that belongs to the Bethe subalgebra.
The existence of this subalgebra lead to the integrability of the 
model when the expansion of the
transfer matrix as a series provides a sufficient number of operators in 
involution. In the following 
we consider that this number is sufficient.   

After proving integrability of the model, the next step is to 
find the eigenvalues and eigenvectors of this Bethe subalgebra. 
It depends on the choice of the boundary matrices. 
Focusing on 
diagonalizable boundary matrices, two main cases can be 
distinguished: $K^{+}(u)$ and $K^-(v)$ are diagonalizable in the 
same basis; or not. 

Very few is known in the latter case, apart 
from two recent approaches developped for $XXZ$ spin chain 
and that does not rely 
on (nested) algebraic Bethe ansatz \cite{FadTak,KuSkly,KR}:  
in \cite{BK}, the reflection equation is 
replace by a deformed Onsager algebra (which may be another 
presentation of the reflection algebra); and in \cite{TQ}, 
eigenvalues are computed using generalized TQ relations when  $K^-(u)$ 
and $K^+(u)$ obey some relations, or when the deformation parameter 
is root of unity. 

The first case can be divided into two 
sub-families, depending whether $(i)$ the diagonalization matrix is a 
constant or $(ii)$ depends on the spectral parameter. Again, 
in the case $(ii)$, only some 
results are known from the gauge transformation 
construction of \cite{gauge,GM,YZ}, that allows to relate non-diagonal 
solutions to diagonal ones via a Face-Vertex correspondence. 
The case $(i)$ is the one studied by analytical Bethe ansatz 
\cite{reshe} and corresponds to diagonal matrices
\ben
K^-(u) &=& diag(\underbrace{k^-(u),\dots,k^-(u)}_{a_{-}},
\underbrace{\bar{k}^-(u),\dots,\bar{k}^-(u)}_{\fm+\fn-a_{-}}),\\
K^+(u) &=& diag(\underbrace{k^+(u),\dots,k^+(u)}_{a_{+}},
\underbrace{\bar{k}^+(u),\dots,\bar{k}^+(u)}_{\fm+\fn-a_{+}})\,.
\een
Indeed, 
using this ansatz, eigenvalues of the transfer matrix can be computed for all 
(open or closed) chains based on $gl(\fn)$ and 
$gl(\fm|\fn)$ 
(super)algebras and their deformation, and with arbitrary representations 
on each sites \cite{ACDFR,ACDFR2,RS}. It is in general believed that a 
Nested (algebraic) Bethe Ansatz (NBA)  can provide the eigenvectors of 
the corresponding models. This was shown in a unified way in \cite{BR1} 
for closed spin chains. 
 
We present here the open spin chains case.  We will show that the 
standard NBA approach does \underline{not} work in the general case. 
Keeping a general diagonal solution for $K^-(u)$, one needs to take 
$a_{+}=0$ or $a_{+}=1$ to perform a complete nested algebraic Bethe 
ansatz. In this case, the couple 
$(K^-(u)\,,\,K^+(u))$ will be called a \textbf{NABA couple}.
When one studies an open spin chain 
possessing  a couple of diagonal matrices $(K^-(u)\,,\,K^+(u))$ that is not 
of type NABA, one can start the first step of NBA approach, but 
then needs to switch (and end) the calculation with an 
analytical Bethe ansatz, as it has been done in e.g. \cite{DVGR,GM,YZ}.  

In the present paper, we focus on NABA couples. 
Performing NBA, we compute the Bethe ansatz equations, the eigenvalues 
and the eigenvectors of the corresponding transfer
matrix and show where the 
constraint $a_{+}=0$ or $a_{+}=1$ is needed in the calculation.
 Our presentation consider 
\textbf{universal transfer matrices}  in the sense that the 
calculation applies to transfer matrices based on $gl(\fn)$ and 
$gl(\fm|\fn)$ 
algebras and their deformation, with any finite dimensional irreducible 
representations of the monodromy matrix. In particular, it encompasses 
the preview results obtained for
 fundamental representations \cite{DVGR,GM,YFH}.  

In addition to the derivation of the Bethe ansatz equations and 
transfer matrix spectrum, our main result is the explicit 
construction of the Bethe vectors. This is reflected in e.g. the trace 
formula (see theorem \ref{theo:traceForm} at the end of the paper).

The plan of the paper is as follows. 
In section \ref{sec:graded-aux}, we introduce the different notations
 and $R$-matrices we use in the paper. Then, in section \ref{sec:alg}, 
we present, using the FRT \cite{FRT}
formalism, the algebras concerned with our approach. They are 
generalizations of loop algebras (quantum algebras or Yangians, and 
their graded versions) and 
noted $\cA_{\fm|\fn}$. They contain as subalgebras the
reflection algebras, noted $\fD_{\fm|\fn}$.
We also construct mappings
$$
\fD_{\fm|\fn} \to \fD_{\fm-1|\fn}  \to \dots \to  \fD_{1|1} 
\mbox{ or }\fD_{2} 
$$
that are needed for the nesting.
In section \ref{sec:hw},  we present
 the finite dimensional irreducible representations of 
$\cA_{\fm|\fn}$ and we compute the form of $T^{-1}(u)$. 
We also construct the representations of $\fD_{\fm|\fn}$ from 
the $\cA_{\fm|\fn}$ ones.
In section \ref{sec:ABA}, as a warm up, we recall the algebraic Bethe ansatz, 
which deals with spin chains based on $gl(2)$, $gl(1|1)$ algebras and 
their quantum deformations. 
Then, in section \ref{sec:NBA}, we perform the nested Bethe ansatz in a very 
detailled and pedestrian way 
and up to the end. Finally, in section \ref{sec:betheV}, we study the Bethe 
vectors that have been constructed 
in the prevous section, showing connection with a \textbf{trace formula}. 
As a conclusion, we discuss our results
and present some possible applications or extensions of our work.

%%%%%%%%%%%%%%%%%%%%%%%%%%%%%%%%%%%%%%%%%%%%%%%%%%%
\section{Notations\label{sec:graded-aux}}

\subsection{Graded auxiliary spaces} 
We use the so-called auxiliary space framework.
In this formalism, one deals with multiple tensor product of 
vectorial spaces 
$\cV \otimes \dots \otimes \cV$, and operators (defining an algebra 
$\cA$) therein.
For any matrix valued operator, 
$A:= \sum_{ij} E_{ij} \otimes a_{ij} \in End(\cV) \otimes \cA\,,
$ we set
\ben
A_k:=\sum_{ij} \II^{\otimes (k-1)}\otimes E_{ij} \otimes  
\II^{\otimes (m-k)}
\otimes  a_{ij}  \in End(\cV^{\otimes m}) \otimes \cA\,,
\ 1 \leq k \leq m \,,
\een
where $E_{ij}$ are elementary matrices, with 1 at
position $(i,j)$ and 0 elsewhere.
The notation is valid for complex matrices, taking $\cA:= \CC$ 
and using the isomorphism $End(\cV) \otimes \CC \sim End(\cV)$.

We will work on $\ZZ_{2}$-graded spaces $\CC^{\fm\vert \fn}$.
The elementary $\CC^{\fm\vert \fn}$ column vectors $e_{i}$ (with 1 at 
position $i$ and 0 elsewhere) and elementary 
$End(\CC^{\fm\vert\fn})$ matrices $E_{ij}$ have grade: 
\begin{equation}
[e_{i}]=[i] \mb{and} [E_{ij}]=[i]+[j]. 
\end{equation}
This grading is also extended to the superalgebras we deal with, see 
section \ref{FRT-for} below.
The tensor product is graded accordingly:
\begin{equation}
(a_{ij} \otimes a_{kl})(a_{pq} \otimes a_{rs}) =
(-1)^{([k]+[l])([p]+[q])}(a_{ij}a_{pq}\otimes a_{kl}a_{rs})\,.
\end{equation}
The transposition $(.)^t$ and trace $str(.)$ operators are also graded:
\ben
A^t = \sum_{i,j=1}^{\fm+\fn} (-1)^{[j]+[j][i]}\,E_{ji} \otimes \,a_{ij}
\,,\quad str A = \sum_{i=1}^{\fm+\fn} (-1)^{[i]}\,a_{ii}
 \mb{for} A = \sum_{i,j=1}^{\fm+\fn}\,E_{ij}\,\otimes a_{ij}.
\een
To simplify the presentation we work with the
\textit{distinguished}
$\ZZ_{2}$-grade defined by:
\begin{equation}
[i]=\left\{ \begin{array}{ll}  0\,, & 1 \leq i \leq \fm\,,\\
			       1\,, & \fm+1\leq i \leq \fm+\fn\,.
  \end{array}\right.
\end{equation}
Simplification in the expressions follows from
 the rule $[i][j]=[i]$ when $i\leq j\,$,
which is valid only for the distinguished grade. 
The non graded case is recovered setting $\fn=0$ and $[k]=0$.

\subsection{Spectral parameters transformations\label{SPECPARA}}
For spectral parameter $u$ we use the following notations:
\beq\begin{array}{ll}
\inv(u)=\begin{cases}
 -u  &\ \mbox{ for $\cY(\fm|\fn)$}\\[0.2ex]
 \frac{1}{u} &\ \mbox{ for $\wh\cU_q(\fm|\fn)$}
\end{cases}&;\quad 
\wt u =
\begin{cases}
u-\frac{(\fm-\fn) \hbar}{2} &\ \mbox{ for $\cY(\fm|\fn)$}\\[0.2ex]
u\,q^{\frac{(\fm-\fn)}{2}} &\ \mbox{ for $\wh\cU_q(\fm|\fn)$}
\end{cases};
\nonumber\\[4.2ex]
u^{(k)}=
\begin{cases}
u+\frac{\hbar}{2}(-1)^{[k]} &\ \mbox{ for $\cY(\fm|\fn)$}\\[0.2ex]
u\,q^{-\frac{1}{2}+2[k]}&\ \mbox{ for $\wh\cU_q(\fm|\fn)$}
\end{cases}&;\quad 
u^{\wb{(k)}}=
\begin{cases}
u-\frac{\hbar}{2}(-1)^{[k]} &\ \mbox{ for $\cY(\fm|\fn)$}\\[0.2ex]
u\,q^{\frac{1}{2}-[k]}&\ \mbox{ for $\wh\cU_q(\fm|\fn)$}
\end{cases};
\nonumber\\[4.2ex]
u^{(k\dots l)}=(\dots (u^{(k)})^{(k+1)}\dots)^{(l)}\,
.
\end{array}
\eeq

\subsection{$R$-matrices}

In what follows, we will deal with different types of 
matrices $R \in End(\cV) \otimes End(\cV)$, 
all obeying (graded) Yang-Baxter equation 
 (writen in auxiliary space 
$End(\cV) \otimes End(\cV) \otimes End(\cV)$):
\ben
R_{12}(u_{1},u_{2})\ R_{13}(u_{1},u_{3})\
R_{23}(u_{2},u_{3})&=&R_{23}(u_{2},u_{3})\ R_{13}(u_{1},u_{3})\
R_{12}(u_{1},u_{2}).
  \label{YBE}
\een
The $R$-matrix satisfy unitarity relation,
\ben
R_{12}(u,v) R_{21}(v,u) &=&\zeta(u,v) \,\II \otimes \II\,,
\label{Unit}
\een
and crossing unitarity (see (\ref{RUBAR}) below for the definition of 
$\bar{R}$), 
\ben
  R^{t_1}_{12}(u, v) M_1 \bar{R}^{t_2}_{12}(\inv(\wt{v}),\inv(\wt u)) 
M_1^{-1} 
&=&\wb \zeta(u,v) \,\II \otimes \II\,,
  \label{CrossU}
\een
where $\zeta(u,v)$ and $\wb \zeta(u,v)$ are $\CC$-functions depending 
on the model under consideration, $M$ is a  $\CC$-valued matrix defined in  
appendix \ref{appendix1} for each model and $t_a$ is the transposition in 
the 
auxiliary space $a$.
All the $R$-matrices used here also obey the parity relation:
\begin{equation}
R_{12}(u,v)^{t_{1}t_{2}}\ =\ R_{21}(u,v).
\end{equation}
To each $R$-matrix, one associates an algebra $\cA_{\fm|\fn}$
 using the FRT formalism. 
Below, we focus on infinite
dimensional associative algebras based on $gl(\fn)$ and $gl(\fm|\fn)$ 
Lie (super)algebras and their $q$-deformation.
We note these algebras $\cA_{\fn}= Y(\fn)$ or $\wh\cU_{q}(\fn)$ and 
$\cA_{\fm|\fn}= \cY(\fm|\fn)$ or $\wh\cU_{q}(\fm|\fn)$.  
We will write also $\cA_{\fm|0}=\cA_{\fm}$.
We will encompass all $R$-matrices of these algebras writing:
\ben
R_{12}(u,v) &=& \fb(u,v)\,\II \otimes \II
+ \sum_{i,j=1}^{{\fm+\fn}} \fw_{ij}(u,v)  E_{ij}
\otimes E_{ji}.
\label{RU}
\een
All functions are defined in appendix \ref{app:fct} for each cases (one 
can refer to \cite{BR1} for details and references).
To define the reflection equation, we need another $R$-matrix:
\ben
\bar{R}_{12}(u,v) &=& R_{12}(u,\inv(v))= \bar \fb(u,v)\II \otimes \II
+ \sum_{i,j=1}^{{\fm+\fn}} \bar \fw_{ij}(u,v)  E_{ij}
\otimes E_{ji}
\label{RUBAR}
\een
From (\ref{YBE}) we can deduce the relation between these two $R$ 
matrices: 
 \ben
R_{12}(u_{1},u_{2})\ \bar{R}_{13}(u_{1},u_{3})\
 \bar{R}_{23}(u_{2},u_{3})&=& \bar{R}_{23}(u_{2},u_{3})\ 
  \bar{R}_{13}(u_{1},u_{3})\
 R_{12}(u_{1},u_{2}).
  \label{YBEB}
\een
We will also use `reduced' $R$-matrices $R^{(k)}(u)$, deduced from 
$R(u)$ by suppressing all the terms containing indices $j$ with $j < 
k$:
\ben
R_{12}^{(k)}(u,v) &=& \left(\II^{(k)} \otimes \II^{(k)}\right)
R_{12}(u,v)\left(\II^{(k)} \otimes \II^{(k)}\right)
\nonu
&=& 
\fb(u,v)\,\II^{(k)} \otimes \II^{(k)}
+ \sum_{i,j=k}^{{\fm+\fn}} \fw_{ij}(u,v)  E_{ij}
\otimes E_{ji}\,,
\label{RUR}\\
\mbox{where} &&\II^{(k)}=\sum_{i=k}^{\fm+\fn}E_{ii}\,,\quad\forall\,k \,.
\label{eq:Ik}
\een
$R_{12}^{(k)}(u,v)$ 
corresponds to the $R$-matrix of\footnote{We will write, for a 
generic $k$, ${\fm-k|\fn}$, keeping 
in mind that one should write ${0|\fn-(k-\fm)}$ when $k>\fm$.} 
$\cA_{\fm+1-k |\fn}$.
We will also use: 
\ben
R_{12}^{(k,p)}(u,v)&=&
\left(\II^{(k)} \otimes \II^{(p)}\right) R_{12}(u,v)
\left(\II^{(k)} \otimes \II^{(p)}\right) \nonu
 &=& \fb(u,v)\,\II^{(k)} \otimes \II^{(p)}
+ \sum_{i,j=max(k,p)}^{{\fm+\fn}} \fw_{ij}(u,v)  E_{ij}
\otimes E_{ji}.
\label{eq:Rpk}
\een
Note that $R_{12}^{(k,k)}(u,v)=R_{12}^{(k)}(u,v)$.
We define the 'normalized reduced' $R$-matrices: 
\ben
\RR_{12}^{(k,p)}(u,v)=
\frac{1}{\fa_{p}(u,v)}\,R_{12}^{(k,p)}(u,v) 
\mb{with} 
\RR^{(k,k)}_{12}(u,v)\,\RR^{(k,k)}_{21}(v,u)=\II^{(k)}\otimes\II^{(k)}.
\label{Runit}
\een

\section{Algebraic structures\label{sec:alg}}

\subsection{FRT formalism \label{FRT-for}} 

The FRT (or RTT) relations \cite{D1,D2,FRT} allow us to 
generate all the relations between the generators
of the graded unital associative algebra $\cA_{\fm|\fn}$.
We gather the $\cA_{\fm|\fn}$ generators  into a 
$(\fm+\fn) \times (\fm+\fn)$  matrix 
acting in an auxiliary space $\cV=\CC^{\fm|\fn}$ whose entries 
are formal series of a complex parameter $u$:
\beq
T(u)= \sum_{i,j=1}^{\fm+\fn}  E_{ij} \otimes t_{ij}(u)
\in  End(\cV) \otimes \cA[[u,u^{-1}]] . 
\eeq
Since the auxiliary space $End(\CC^{\fm|\fn})$ is interpreted as a 
representation of $\cA_{\fm|\fn}$, 
the $\ZZ_{2}$-grading of $\cA_{\fm|\fn}$ must correspond to the one 
defined on $End(\CC^{\fm|\fn})$ matrices (see section 
\ref{sec:graded-aux}). 
Hence, the generator $t_{ij}(u)$ has grade $[i]+[j]$, so that the 
monodromy matrix $T(u)$ is globally even. As for matrices, the tensor 
product 
of algebras will be graded, as well as between algebras and matrices, 
\beq
\big(E_{ij} \otimes t_{ij}(u)\big)\,\big(E_{kl} \otimes 
t_{kl}(v)\big)=
(-1)^{([i]+[j])([k]+[l])}\,E_{ij}\,E_{kl} 
\otimes t_{ij}(u)\,t_{kl}(v)\,.
\eeq
The `real' generators $t_{ij}^{(n)}$ of $\cA_{\fm|\fn}$ appear upon 
expansion of 
$t_{ij}(u)$ in $u$. For the (super) Yangians $\cY(\fn)$ and 
$\cY(\fm|\fn)$, 
$t_{ij}(u)$ is a series in $u^{-1}$:
\begin{equation}
t_{ij}(u)=\sum_{n=0}^\infty t_{ij}^{(n)} u^{-n} \mb{with}
t_{ij}^{(0)}=\delta_{ij}\,.
\label{eq:expT}
\end{equation}
In the quantum affine (super)algebra \cite{Jim,ReSem} without central charge case, 
a complete description of
the algebras requires the introduction of two matrices $L^\pm(u)$, 
% \ben
% L^\pm(u)&=&\sum_{i,j=1}^{\fm+\fn}  E_{ij} \otimes L^\pm_{ij}(u) 
% =\sum_{i,j=1}^{\fm+\fn}  E_{ij} \otimes \sum_{n=0}^\infty 
% L^{\pm (n)}_{ij} u^{\pm 2n},
% \een
% with  relations: 
% \ben
% L^{+ (0)}_{ii}L^{- (0)}_{ii}=1\,,\,\forall\,i &\mbox{and}&
% L^{+(0)}_{ij}=0=L^{- (0)}_{ji}\,,\ i<j,\\
% R_{12}(u,v)\ L^\pm_{1}(u)\ L^\pm_{2}(v) 
% &=&L^\pm_{2}(v)\ L^\pm_{1}(u)\ R_{12}(u,v)\,,\\
% R_{12}(u,v)\ L^\mp_{1}(u)\ L^\pm_{2}(v)
% &=&L^\pm_{2}(v)\ L^\mp_{1}(u)\ R_{12}(u,v)\,.
%   \label{RLL}
% \een
However, in the context of evaluation representations it is 
sufficient 
to consider only $T(u)=L^+(u)$ to 
construct a transfer matrix. Indeed, in an
evaluation representation, the choices  $T(u)=L^-(u)$ or  
$T(u)=L^+(u)-L^-(u)$ lead to the
same operator up to a multiplication function. 
Then, the RTT relations take the form:
\beq
  R_{12}(u,v)\ T_{1}(u)\ T_{2}(v)=T_{2}(v)\ T_{1}(u)\ R_{12}(u,v)\,. 
  \label{RTT}
\eeq
$\cA_{\fm|\fn}$ has the following antimorphisms:
\begin{equation}\begin{array}{l}
\mb{Matrix inversion} inv :\quad  T(u)\,\to\,T^{-1}(u)=
\sum_{i,j=1}^{\fm+\fn} E_{ij}\otimes t'_{ij}(u)\,,\\[1.2ex]
\mb{Matrix transposition}  t :\quad  
T(u)\,\to\,T^{t}(u)=\sum_{i,j=1}^{\fm+\fn}(E_{ij})^t \otimes 
t_{ij}(u)\,,\\[1.2ex]
\mb{Spectral parameter inversion} \inv :\quad  
T(u)\,\to\,T(\inv(u))\,.
\end{array}
\end{equation}
$\cA_{\fm|\fn}$ has a Hopf algebra structure, with
coproduct  
\beq
\Delta(T(u)) =
T(u)\dot{\otimes}T(u) = \sum_{i,j,k=1}^{\fm+\fn} 
(-1)^{([k]+[i])([k]+[j])}\,
E_{ij}\otimes t_{ik}(u)\otimes t_{kj}(u)\,.
\eeq
More generally, one defines recursively for $L\geq 2$, the
algebra homomorphism
\begin{equation}
\Delta^{(L+1)}=(\id^{\otimes (L-1)}\otimes \Delta)\circ\Delta^{(L)}\
\mb{with} \Delta^{(2)}=\Delta \mb{and}\Delta^{(1)}=\id\,. 
\end{equation}

\subsection{Reflection algebra and $K(u)$ matrices} 
The  $\cA_{\fm|\fn}$ algebra is enough to 
construct a transfer matrix leading to
periodic spin chain models.
In the context of open spin chains, one needs another algebra, 
the reflection algebra
$\fD_{\fm|\fn}$, which turns out to be a subalgebra of $\cA_{\fm|\fn}$. 
Indeed, physically, one can interpret the 
 FRT relation as encoding the interaction between the spins of the 
chain. Hence, it is the only relation needed to describe a periodic 
chain. On the other hand, in the case of open chain, the interaction with the 
boundaries has to be taken into account. 
Following the seminal paper of Sklyanin \cite{S}, 
we construct the reflection algebra and the 
dual reflection equation
 for the boundary scalar matrices $K^-(u)$ and $K^+(u)$. 
We first define the matrix $K^-(u)$ to be the solution of the 
reflection equation:
\beq
  R_{12}(u_{1},u_{2}) K^-_{1}(u_{1})
  \bar{R}_{21}(u_{1},u_{2}) K^-_{2}(u_{2})=
  K^-_{2}(u_{2})\bar{R}_{12}(u_{1},u_{2})
  K^-_{1}(u_{1})R_{21}(u_{1},u_{2}) .
  \label{ERK}
\eeq
Depending on the type of $R$-matrix one considers, solutions to the 
reflection equation have been classified: see \cite{AACDFR} for the 
Yangian 
 and  super-Yangian cases; in the other cases, partial 
classifications have been obtained in e.g. \cite{G,DVGR}. 
In all cases, 
 diagonal solutions of the reflection equations are known. They 
take the form (up to normalisation),
\ben
K^-(u)=\begin{cases}
\mbox{ diag}\,(\underbrace{u-c_-,\dots,u-c_-}_a,
\underbrace{-u-c_-,\dots,-u-c_-}_{\fm+\fn-a}) 
&\mb{for $\cY(\fm|\fn)$,}\\
\mbox{ diag}\,(\underbrace{u^2-c_-^2,\dots,u^2-c_-^2}_a,
\underbrace{u^{-2}-c_-^2,\dots,u^{-2}-c_-^2}_{\fm+\fn-a}) 
&\mb{for $\wh\cU_q(\fm|\fn)$,}
\end{cases}
\label{eq:Kdiag}
\een
where $c_{-}$ is a free complex parameter and $a$ is an integer such 
that $0\leq a\leq \fm+\fn$.
From this $K^-(u)$ matrix and the monodromy matrix $T(u)$ for closed 
spin chains, we can construct the monodromy matrix of open 
spin chain:
\ben
D(u)&=&T(u)\,K^-(u)\,T^{-1}(\inv(u)) 
\ =\ \sum_{i,j=1}^{\fm+\fn}  E_{ij}\otimes d_{ij}(u), 
\label{D-from-T}\\
d_{ij}(u) &=& \sum_{a=1}^{\fm+\fn} (-1)^{([i]+[a])([a]+[j])}\,
\kappa_a(u)\,t_{ia}(u)\,t'_{aj}(\inv(u)).
\een
From (\ref{RTT})  and  (\ref{ERK}), we can prove that $D(u)$ also
satisfies the reflection equation: 
\beq
R_{12}(u_{1},u_{2}) D_{1}(u_{1})
\bar{R}_{21}(u_{1},u_{2}) D_{2}(u_{2})=
D_{2}(u_{2})\bar{R}_{12}(u_{1},u_{2})
D_{1}(u_{1})R_{21}(u_{1},u_{2}). 
\label{ER}
\eeq
This relation defines the reflection algebra $\fD_{\fm|\fn}$.
The algebra $\fD_{\fm|\fn}$ is a left coideal \cite{MoRa} of the 
algebra $\cA_{\fm|\fn}$ with coproduct:
\ben
\Delta D_{[2]}(u)=T_{[1]}(u)\,D_{[2]}(u)\,T_{[1]}^{-1}(\inv(u))
\in End(\cV)\otimes\cA_{\fm|\fn}\otimes\fD_{\fm|\fn} 
\label{eq:deltaD}
\een
where $[i]$ labels the two copies of $\cA_{\fm|\fn}$. 
This expression allows to increase the number of sites for an open 
spin chain in the same way one does for periodic ones: one acts on 
the monodromy matrix with the coproduct and then represents the new 
copy of algebra on the new `site'.

We also need a dual equation to construct transfer matrices in 
involution:
\begin{eqnarray}
&&R_{12}(u_{2},u_{1}) (K^+_{1}(u_{1}))^{t_1}
M_1^{-1} \bar{R}_{21}(\inv(\wt u_{1}),\inv(\wt u_{2})) M_1 
(K^+_{2}(u_{2}))^{t_2}=\nonu
&& 
 (K^+_{2}(u_{2}))^{t_2} M_1 \bar{R}_{12}(\inv(\wt u_{1}),\inv(\wt 
u_{2})) M_1^{-1} 
  (K^+_{1}(u_{1}))^{t_1}R_{21}(u_{2},u_{1}). 
  \label{ERD}
\end{eqnarray}
where $M$ is given in appendix \ref{appendix1}. From the property
\beq
R_{12}(u_{1}\,,\,u_{2})\,M_{1}\,M_{2} = M_{1}\,M_{2}\,
R_{12}(u_{1}\,,\,u_{2})\,,
\eeq
one can construct solutions to the dual reflection equation using 
$K^-(u)$ solutions:
\ben
(K^+(u))^t=M \,K^-(\inv(\wt u))\,.
\een
With $D(u)$ and $K^+(u)$ we construct the transfer matrix: 
\beq
d(u) = str(K^+(u) D(u)). 
\eeq
The reflection equation and its dual form ensure the commutation 
relation
$[d(u),d(v)] = 0$. 
Thus, $d(u)$ generates (via an expansion in $u$) a set of $L$ (the 
number of sites)
independent integrals of motion (or charges) in involution which ensure 
integrability of the model. 

\subsubsection{Commutation relations of $\fD_{\fm|\fn}$}

Projecting (\ref{ER})
on the $E_{ij}\otimes E_{kl}$ basis we get the commutation relations
for  $\fD_{\fm|\fn}$:
\ben
&&[d_{ij}(u),d_{kl}(v)\}\ =\ 
-\delta_{kj}\sum_{a=1}^{\fm+\fn}\frac{\bar \fw_{ka}(u,v)}{\bar 
\fb(u,v)}\,(-1)^{([i]+[l])([a]+[k])}\,
d_{ia}(u)d_{al}(v)
\nonu
&&\qquad
+\delta_{il}\sum_{a=1}^{\fm+\fn}\frac{\bar \fw_{ia}(u,v)}{\bar 
\fb(u,v)}\,(-1)^{[l]+[a][k]+[a][j]+[k][j]}\,
d_{ka}(v)d_{aj}(u)\nonumber \\
&&\qquad
-\delta_{ij}\frac{\fw_{ik}(u,v)}{\fb(u,v)}\,\sum_{a=1}^{\fm+\fn}
\frac{\bar \fw_{ia}(u,v)}{\bar \fb(u,v)}\,(-1)^{([a]+[k])([k]+[l])}\,
d_{ka}(u)d_{al}(v)\nonumber \\
&&\qquad
+\delta_{ij}\frac{\fw_{il}(u,v)}{\fb(u,v)}\,\sum_{a=1}^{\fm+\fn}
\frac{\bar \fw_{ia}(u,v)}{\bar \fb(u,v)}\,(-1)^{([a]+[l])([k]+[l])}\,
d_{ka}(v)d_{al}(u)
\nonumber \\
&&\qquad
-\frac{\fw_{ik}(u,v)}{\fb(u,v)}\,(-1)^{([i]+[k])([k]+[l])}\,d_{kj}(u)d_{il}(v)
+\frac{\fw_{jl}(u,v)}{\fb(u,v)}\,(-1)^{([i]+[l])([k]+[l])}\,d_{kj}(v)d_{il}(u)
\qquad\quad
\label{RCP}
\een
where $[x\,,\,y\}= x\,y-(-1)^{[x][y]}\,y\,x$ is the graded commutator.

\subsection{Embeddings of $\fD_{\fm|\fn}$ algebras}
The algebraic cornerstone for the nested Bethe ansatz is a 
recursion relation on
 the $\fD_{m|n}$ algebraic structure. In this section we present a 
 coset construction for 
$\fD_{m|n}$ algebras (see theorem \ref{theo:embedding}), that extends 
to the coideal property (see lemma \ref{lem-1} and theorem \ref{embedding-co}).

\begin{theorem}
\label{theo:embedding}
For $k=1,2,\ldots,\fm+\fn-1$, let $F^{(k)}$ be a linear combination of 
elements $d_{i_1j_1}(u_1)\dots d_{i_lj_l}(u_l)$
with all indices $i_{p},\,j_{p}>k-1$, and let
$\cI_{k}$ be the left ideal generated by 
$d_{ij}(u)$ for $i>j$ {and} $j<k$.
Then,
we have the following properties:
\ben
 d_{ij}(u)\,F^{(k)}&\equiv&0 \mb{mod $\cI_{k}$}\,,\mb{for} i>j \mb{and} 
j<k,\\
\null [d_{ii}(u)\,,F^{(k)}] &\equiv&0 \mb{mod $\cI_{k}$}\,, \mb{for} i<k.
\label{com-LC}
\een
Using the functions $\psi_{j}$ given in (\ref{eq:psij}), we introduce the 
generators:
\ben
\wh D^{(k)}(u)&=&
\sum_{i,j=k}^{\fm+\fn} E_{ij} \otimes d^{(k)}_{ij}(u), \\
d^{(k)}_{ij}(u)&=&d_{ij}(u^{(1\dots 
k-1)})-\delta_{ij}\sum_{a=1}^{k-1} 
\,q^{2(k-1-a)-4\sum_{l=a+1}^{k-1}[l]}\,\psi_a(u^{(a)}) \,
d_{aa}(u^{(1\dots k - 1)})\,.
\label{TF-D}
\een
They 
satisfy in $\fD_{\fm|\fn}/\cI_{k}$ the reflection equation for 
$\fD_{\fm-k+1|\fn}$:
\beq
R^{(k)}_{12}(u_{1},u_{2}) \wh D^{(k)}_{1}(u_{1})
\bar{R}^{(k)}_{21}(u_{1},u_{2})\wh D^{(k)}_{2}(u_{2})
\equiv
\wh D^{(k)}_{2}(u_{2}) \bar{R}^{(k)}_{12}(u_{1},u_{2})
\wh D^{(k)}_{1}(u_{1}) R^{(k)}_{21}(u_{1},u_{2})   \ \mbox{ mod } \cI_{k}\,.
\label{ERbis}
\eeq
\end{theorem}
\prf
We first prove relation (\ref{com-LC}) for $k=2$, the case $k=1$ being 
trivially satisfied. A direct calculation from the commutation 
relations (\ref{RCP}) of $\fD_{\fm|\fn}$  leads to (for $i,j,l>1$): 
\ben
d_{j1}(u)d_{11}(v)
% &=& \frac{\fa_1(u,v) \bar \fb(u,v)}{\fb(u,v) \bar 
% \fa_1(u,v)}d_{11}(v)d_{j1}(u) -\frac{ \fw_{j1}(u,v) \bar 
% \fb(u,v)}{\fb(u,v) \bar \fa_1(u,v)}d_{11}(u)d_{j1}(v) \nonu
% &&-\sum_{a=2}^{\fm+\fn}\frac{\bar \fw_{1a}(u,v)}{\bar 
% \fa_1(u,v)}(-1)^{([j]+[1])([a]+[1])}\,d_{ja}(v)d_{a1}(u),
% \\
&\equiv& 0\mb{mod $\cI_{2}$}
\\
{[d_{ij}(u),d_{l1}(v)\}}
% &=&
% -\frac{\fw_{il}(u,v)}{\fb(u,v)}\,(-1)^{([i]+[l])([l]+[1])}\,d_{lj}(u)d_{i1}(v)
% \nonu
% &&+\frac{\fw_{j1}(u,v)}{\fb(u,v)}\,(-1)^{([i]+[1])([l]+[1])}\,d_{lj}(v)d_{i1}(u)
% \nonu
% &&-\delta_{lj}\sum_{a=1}^{\fm+\fn}\frac{\bar \fw_{la}(u,v)}{\bar 
% \fb(u,v)}\,(-1)^{([i]+[1])([a]+[l])}\,
% d_{ia}(u)d_{a1}(v)
% \nonu
% &&-\delta_{ij}\frac{\fw_{il}(u,v)}{\fb(u,v)}\,\sum_{a=1}^{\fm+\fn}
% \frac{\bar \fw_{ia}(u,v)}{\bar \fb(u,v)}
% \,(-1)^{([j]+[a])([1]+[l])}\,d_{la}(u)d_{a1}(v)
% \nonu
% &&+\delta_{ij}\frac{\fw_{i1}(u,v)}{\fb(u,v)}\,\sum_{a=1}^{\fm+\fn}
% \frac{\bar \fw_{ia}(u,v)}{\bar \fb(u,v)}\,(-1)^{([a]+[1])([1]+[l])}\,
% d_{la}(v)d_{a1}(u),
% \\
&\equiv& 0\mb{mod $\cI_{2}$}
\\
{[d_{ij}(u),d_{11}(v)]}
% -\frac{\fw_{i1}(u,v)}{\fb(u,v)}d_{1j}(u)d_{i1}(v)
% +\frac{\fw_{jl}(u,v)}{\fb(u,v)}d_{1j}(v)d_{i1}(u)
% \nonumber \\
% &&\qquad\qquad
% -\delta_{ij}\frac{\fw_{i1}(u,v)}{\fb(u,v)}\,\sum_{a=2}^{\fm+\fn}
% \frac{\bar \fw_{ia}(u,v)}{\bar \fb(u,v)}
% d_{1a}(u)d_{a1}(v)
% +\delta_{ij}\frac{\fw_{i1}(u,v)}{\fb(u,v)}\,\sum_{a=2}^{\fm+\fn}
% \frac{\bar \fw_{ia}(u,v)}{\bar \fb(u,v)}
% d_{1a}(v)d_{a1}(u)\nonu
% &&\qquad\qquad-\delta_{ij}\frac{\fw_{i1}(u,v)\bar \fw_{i1}(u,v)}{\fb(u,v)\bar 
% \fb(u,v)}
% [d_{11}(u),d_{11}(v)],
% \nonu
&\equiv& -\delta_{ij}\frac{\fw_{i1}(u,v)\,\bar 
\fw_{i1}(u,v)}{\fb(u,v)\,\bar \fb(u,v)}\,
[d_{11}(u),d_{11}(v)]\mb{mod $\cI_{2}$}.
\\
{[d_{11}(u),d_{11}(v)]}
% =\Big(1+\frac{\fw_{11}(u,v)}{\fb(u,v)}+
% \frac{\bar \fw_{11}(u,v)}{\bar \fb(u,v)}
% +\frac{\bar \fw_{11}(u,v)\,\fw_{1a}(u,v)}{\fb(u,v)\,\bar 
% \fb(u,v)}\Big)^{-1}
%  \times \nonu
% &&\times\Big(-\sum_{a=2}^{\fm+\fn}\frac{\bar \fw_{1a}(u,v)}{\bar \fb(u,v)}
% d_{1a}(u)d_{a1}(v)
% +\sum_{a=2}^{\fm+\fn}\frac{\bar \fw_{1a}(u,v)}{\bar \fb(u,v)}
% d_{1a}(v)d_{a1}(u)
% \nonumber \\
% &&\quad-\frac{\fw_{11}(u,v)}{\fb(u,v)}\,\sum_{a=2}^{\fm+\fn}
% \frac{\bar \fw_{1a}(u,v)}{\bar \fb(u,v)}
% d_{1a}(u)d_{a1}(v)
% +\frac{\fw_{11}(u,v)}{\fb(u,v)}\,\sum_{a=2}^{\fm+\fn}
% \frac{\bar \fw_{1a}(u,v)}{\bar \fb(u,v)}
% d_{1a}(v)d_{a1}(u))\Big),
% \nonu
&\equiv& 0\mb{mod $\cI_{2}$}.
\een
Gathering all these equations, we get relation 
(\ref{com-LC}) for $k=2$.

We now prove relation (\ref{ERbis})
 for $k=1,2$.
For $k=1$, $d^{(1)}_{ij}(u)=d_{ij}(u)$, the ideal $\cI_{1}$ is empty, and we have the starting 
algebra $\fD_{\fm|\fn}$, so that relation (\ref{ERbis}) for $k=1$ just 
corresponds to the 
standard reflection equation.   
For $k=2$, we use the following commutation relations:
\ben
[d_{ij}(u),d_{11}(v)\} &\equiv& 0 \mb{for} i,j>1\mb{and}
[d_{11}(u),d_{11}(v)\} \ \equiv\ 0
\label{eq:truc},\\
d_{j1}(u) &\equiv& 0 \mb{for $j> 1$}
\mb{mod $\cI_{2}$}.
\een
This implies that for $i,j,g,l>1$, we have:
\ben
[d^{(2)}_{ij}(u), 
d^{(2)}_{gl}(v)\}\equiv[d_{ij}(u^{(1)}),d_{gl}(v^{(1)})\}
\mb{mod $\cI_{2}$}.
\een
Hence, it just remains to prove that the relation can be 
re-expressed in terms of $ d^{(1)}_{rs}(u)$, $r,s>1$, only.
For such a purpose, we compute the commutation 
relations between $d_{i1}(u)$ and $d_{1j}(v)$:
\ben
&&d_{i1}(u)\,d_{1j}(v) \equiv (-1)^{[j]([j]+[i])}\,
\frac{\fw_{1j}(u,v)}{\fb(u,v)}
\,d_{11}(v)\,d_{ij}(u)
+\delta_{ij} \,(-1)^{[j]+[1]}
\, \frac{\bar \fw_{i1}(u,v)}{\bar \fb(u,v)}\,
d_{11}(v)\,d_{11}(u)
\ \, \nonu
&&-(-1)^{[1]+[j]\,[i]}
\,\frac{\fw_{i1}(u,v)}{\fb(u,v)}\,d_{11}(u)\,d_{ij}(v)
-\sum_{a=1}^{\fm+\fn}(-1)^{([i]+[j])[a]}\,
\frac{\bar \fw_{1a}(u,v)}{\bar \fb(u,v)}\,
d_{ia}(u)\,d_{aj}(v)
\,.\qquad 
\label{eq:di1d1j}
\een
Using this equation, one can compute\footnote{To obtain this equation, we started from the lhs of 
(\ref{BCcom}), and changed the term $a=1$ according to 
(\ref{eq:di1d1j}).} for any polynomial function $f([a],[b],[c],[d])$: 
\ben
&&\sum_{a=1}^{\fm+\fn}(-1)^{f([i],[l],[a],[g])}\,
\frac{\bar \fw_{ga}(u^{(1)},v^{(1)})}{\bar \fb(u^{(1)},v^{(1)})}
\,d_{ia}(u^{(1)})\,d_{al}(v^{(1)})
 \equiv
\nonu
&&\equiv\qquad \sum_{a=2}^{\fm+\fn}
(-1)^{f([i],[l],[a],[g])}\,
\frac{\bar \fw_{ga}(u,v)}{\bar \fb(u,v)}
\,d_{ia}(u^{(1)})\,d_{al}(v^{(1)})
 \nonu
 &&\qquad-
 (-1)^{[1]+[g]\,[i]+f([i],[l],[1],[g])}\,
\frac{\fw_{i1}(u,v)\,\bar \fw_{g1}(u^{(1)},v^{(1)})}
 { \fb(u,v)\, \bar \fa_1(u^{(1)},v^{(1)})}
 \,d_{11}(u^{(1)})\,d_{il}(v^{(1)}) \nonu
 &&\qquad +
 (-1)^{[l]([l]+[i])+f([i],[l],[1],[g])}\,
\frac{\fw_{1l}(u,v)\,\bar \fw_{g1}(u^{(1)},v^{(1)})}{ 
\fb(u,v)\, \bar \fa_1(u^{(1)},v^{(1)})}
 \,d_{11}(v^{(1)})\,d_{il}(u^{(1)}) \nonu
 &&\qquad+\delta_{il} \, 
 (-1)^{[1]+[i]+f([i],[l],[1],[g])}\,
\frac{\bar \fw_{i1}(u^{(1)},v^{(1)})\,
 \bar \fw_{g1}(u^{(1)},v^{(1)})}
 {\bar \fb(u^{(1)},v^{(1)})\,\bar \fa_1(u^{(1)},v^{(1)})}
 \,d_{11}(u^{(1)})\,d_{11}(v^{(1)})\,.\qquad
 \label{BCcom}
 \een
 Now using (\ref{eq:truc}), we can write the commutation relations 
for 
the new operator $d^{(1)}_{ij}(u)$ with (\ref{RCP}).
We extract the term $d_{i1}(u)\,d_{1j}(v)$ of each sum and using 
(\ref{BCcom})
we are left with only operators $d_{11}(u)$ and $d_{ij}(u)$ with $i,j \neq 1$. 
We use the transformation 
\ben
d^{(2)}_{ij}(u)&=&d^{(1)}_{ij}(u^{(1)})-\delta_{ij}\,\psi_1(u^{(1)})\,d^{(1)}_{11}(u^{(1)}),
\een
and we obtain the desired commutation relation plus some 
unwanted terms.
 It is a straightforward calculation to find
that the unwanted terms cancel for all value of $i,j,g,l \in \{2,\dots 
\fm+\fn\}$.
This proves that $\hat d_{ij}(u)$ has the same commutation relation as 
(\ref{RCP}) 
with a sum starting at $2$ for $i,j>1$. Thus, theorem
\ref{theo:embedding} is proved for $k=2$.
 
For $k>2$, we first remark that one has $\cI_{k}\subset\cI_{k+1}$, 
so that 
one can use the results of step $k$ in the proof of step $k+1$.
Then, the calculation becomes equivalent to the $k=2$ case.
The transformation of $D$ operator for each step is: 
\ben
D^{(k)}(u) &=& \II^{(k)}\,\wh D^{(k-1)}(u)\,\II^{(k)}\,,\ k>1\\
\wh D^{(k)}(u) &=& \II^{(k)}\Big(\wh 
D^{(k-1)}(u^{(k-1)})-\psi_{k-1}(u^{(k-1)})
\,d_{kk}^{(k-1)}(u^{(k-1)})\, \II^{(k)} \Big)\II^{(k)}\,,\ k>1
\label{TFO}\\
D^{(1)}(u) &=& \wh D^{(1)}(u) = \II^{(1)}\, D(u)\,\II^{(1)}  = D(u)
 \een
A direct calculation using identity 
(\ref{eq:id-psi}) gives the form (\ref{TF-D}).
\finprfbis

\begin{lemma}\label{lem-1} 
For $k=1,2,\ldots,\fm+\fn-1$, let $G^{(k)}$ be a linear combination of 
elements,
$$t_{i_1j_1}(u_1)\,t'_{g_1l_1}(\inv{(u_1)})\dots 
t_{i_pj_p}(u_p)\,t'_{g_pl_p}(\inv{(u_p)}),$$
with all indices $i_{r},\,j_{r},\,g_{r},\,l_{r}>k-1$,
and
let $\cJ_{k}$ be the left ideal generated by
$\{t_{ij}(u)\,,\ t'_{ij}(u)\}$ for $i>j \mb{and}  j<k$. 
Then, we have the following properties:
\ben
 t_{ij}(u)\,G^{(k)}&\equiv&0 \mb{mod $\cJ_{k}$}\,
\mb{for} i>j \mb{and} j<k,
\label{com-LC-2deb}
\\
t'_{ij}(u)\,G^{(k)}&\equiv&0 \mb{mod $\cJ_{k}$}\,\mb{for} i>j 
\mb{and} j<k,\\
\null [t_{ii}(u)\,,G^{(k)}] &\equiv&0 \mb{mod $\cJ_{k}$}\mb{for} i<k,\\
\null [t'_{ii}(u)\,,G^{(k)}] &\equiv&0 \mb{mod $\cJ_{k}$}\mb{for} i<k.
\label{com-LC-2fin}
\een
Moreover, the generators, 
\ben
 T^{(k)}(u)&=&
\sum_{i,j=k}^{\fm+\fn} E_{ij} \otimes t^{(k)}_{ij}(u) \mb{and}
(T^{-1})^{(k)}(\inv{(u)}) = \sum_{i,j=k}^{\fm+\fn} E_{ij} \otimes 
t'^{(k)}_{ij}(\inv{(u)}),
\\
t^{(k)}_{ij}(u)&=&t_{ij}(u^{(1\dots k-1)}) \mb{ and}
t'^{(k)}_{ij}(\inv{(u)}) = t'_{ij}(\inv(u^{(1\dots k-1)})),
\label{TF-T-T'}
\een
satisfy in ${\cA}_{\fm|\fn}/\cJ_{k}$ the relation:
\ben
(T^{-1}_2)^{(k)}(\inv{(u)})\,R^{(k)}_{12}(u,\inv{(u)})\,T^{(k)}_1(u)\,
\equiv T^{(k)}_1(u)\,R^{(k)}_{12}(u,\inv{(u)})\,(T^{-1}_2)^{(k)}(\inv{(u)})\, 
\mb{mod $\cJ_{k}$ .}
\label{TRT-k}
\een
\end{lemma}
\prf
As for theorem \ref{theo:embedding}, the case $k=1$ is just the definition 
of the algebra and relations
(\ref{com-LC-2deb})-(\ref{com-LC-2fin}) do not existe.

We prove the case $k=2$, the proof for the other cases being similar. 
From the relation  
\ben
T^{-1}_2(v)\,R_{12}(u,v)\,T_1(u)=T_1(u)\,R_{12}(u,v)\,T^{-1}_2(v)
\een
one gets by projecting on $E_{ij}\otimes E_{gl}$:
\ben
\Big[t'_{ij}(u)\,,t_{gl}(v)\Big\}&=&
\delta_{il}\, \sum_{a=1}^{\fm+\fn}\, (-1)^{([j]+[a])([g]+[a])}\,
 \frac{\fw_{al}(v,u)}{\fb(v,u)}\, 
t_{ga}(v)\,t'_{aj}(u) 
\nonumber \\
&&-\delta_{jg}\, \sum_{a=1}^{\fm+\fn}\, 
(-1)^{([g]+[a])([i]+[l])}\, \frac{\fw_{ga}(v,u)}{\fb(v,u)}\, 
t'_{ia}(u)\,t_{al}(v). 
\label{TRT}
\een
From (\ref{TRT}) we find for $i,j,g,l \neq 1$:
\ben
t'_{i1}(u)\,t_{11}(v)
% &=& \frac{1}{\fa_1(v,u)}\,t_{11}(v)t'_{i1}(u)
% -\sum_{a=2}^{\fm+\fn} \frac{\fw_{ga}(v,u)}{\fa_1(v,u)}
% \, (-1)^{([1]+[a])([i]+[1])}\, 
% t'_{ia}(u)\,t_{a1}(v),
% \nonumber
\ \equiv\ 0 \mb{mod $\cJ_{2}$}
\mb{ ; }
\Big[t'_{ij}(u)\,,t_{g1}(v)\Big\}
% &=&
% -\delta_{jg} \sum_{a=2}^{\fm+\fn} \frac{\fw_{ga}(v,u)}{\fb(v,u)}
% \, (-1)^{([g]+[a])([i]+[1])}\, 
% t'_{ia}(u)\,t_{a1}(v) 
%  \nonu
% &&-\delta_{jg} \frac{\fw_{g1}(v,u)}{\fb(v,u)}
% \, (-1)^{([g]+[1])([i]+[1])}
% \, t'_{i1}(u)\,t_{11}(v), 
&\equiv& 0 \mb{mod $\cJ_{2}$}
\label{eq:toto1}
\\
t_{g1}(v)\,t'_{11}(u)
% &=&\frac{1}{\fa_1(v,u)}t'_{11}(u)t_{g1}(v)
% - \sum_{a=2}^{\fm+\fn} \frac{\fw_{al}(v,u)}{\fa_1(v,u)}
% \, (-1)^{([j]+[a])([g]+[a])}\, 
% t_{ga}(v)\,t'_{a1}(u). \quad
\ \equiv\ 0 \mb{mod $\cJ_{2}$}
\mb{ ; }
\Big[t'_{i1}(u)\,,\,t_{gl}(v)\Big\}
% &=&
% \delta_{il} \sum_{a=2}^{\fm+\fn} \frac{\fw_{al}(v,u)}{\fb(v,u)}
% \, (-1)^{([j]+[a])([g]+[a])}\, 
% t_{ga}(v)\,t'_{a1}(u)
% \nonu
% &&+\delta_{il} \frac{\fw_{1l}(v,u)}{\fb(v,u)}\, 
% t_{g1}(v)\,t'_{11}(u), 
% \nonu
&\equiv& 0 \mb{mod $\cJ_{2}$}
\een
We also need the following commutation relation proved in \cite{BR1}:
\ben
\Big[t_{ij}(u)\,,t_{k1}(v)\Big\} &\equiv&0 \mb{mod $\cJ_{2}$ .}
\een
In the same way, one can compute: 
\ben
\Big[t'_{ij}(u)\,,t'_{k1}(v)\Big\} &\equiv&0 \mb{mod $\cJ_{2}$ .}
\label{eq:toto-fin}
\een
Starting from the left-hand-side of relations 
(\ref{com-LC-2deb})-(\ref{com-LC-2fin}), a recursive use of 
commutation relations (\ref{eq:toto1})-(\ref{eq:toto-fin})
 prove that one gets only terms with $t_{a1}(u)$ and $t'_{a1}(u)$ on 
 the right, so that properties 
 (\ref{com-LC-2deb})-(\ref{com-LC-2fin}) hold for $k=2$.

To prove (\ref{TRT-k}), we start again with relation (\ref{TRT}) with $i,j,g,l \neq 1$
 and extract the first term in the summation:
\ben
&&\Big[t'_{ij}(u)\,,t_{gl}(v)\Big\}\ =\ 
\delta_{il} \sum_{a=2}^{\fm+\fn}\, (-1)^{([j]+[a])([g]+[a])}
\, \frac{\fw_{al}(v,u)}{\fb(v,u)}\, 
t_{ga}(v)\,t'_{aj}(u) 
\nonumber \\
&&\qquad\quad-\delta_{jg} \sum_{a=2}^{\fm+\fn}
\, (-1)^{([g]+[a])([i]+[l])}\, \frac{\fw_{ga}(v,u)}{\fb(v,u)}\, 
t'_{ia}(u)\,t_{al}(v) 
 \nonu
&&\qquad\quad+\delta_{il}\, (-1)^{[j][g]+[1]}
\, \frac{\fw_{1l}(v,u)}{\fb(v,u)}\, t_{g1}(v)\,t'_{1j}(u) 
-\delta_{jg}\, (-1)^{[g]([i]+[l])}
\, \frac{\fw_{g1}(v,u)}{\fb(v,u)}\, t'_{i1}(u)\,t_{1l}(v). 
\qquad\quad
\een
Inserting in this equation the relations (valid modulo $\cJ_2$):
\beq
 t'_{i1}(u)\,t_{1l}(v) \equiv -\sum_{a=2}^{\fm+\fn} 
 (-1)^{[a]([i]+[l])}\,\frac{\fw_{1a}(v,u)}{\fa_1(v,u)}\, 
t'_{ia}(u)\,t_{al}(v) 
+\delta_{il} \,
\frac{\fw_{1l}(v,u)}{\fa_1(v,u)}\,t'_{11}(u)\,t_{11}(v),  
\eeq
\ben
(-1)^{[1]+[i][j]}\,t_{i1}(v)\,t'_{1j}(u) &\equiv& 
-\sum_{a=2}^{\fm+\fn} (-1)^{([a]+[j])([a]+[i])}
\,\frac{\fw_{a1}(v,u)}{\fa_1(v,u)}\, 
t_{ia}(v)\,t'_{aj}(u)
\nonumber \\
&&
 +\delta_{ji} \,
\frac{\fw_{1l}(v,u)}{\fa_1(v,u)}\,t'_{11}(u)\,t_{11}(v) ,
\\
\Big[t'_{11}(u)\,,t_{11}(v)\Big]&\equiv&0,
\een
and making the transformation $u \to \inv{(u^{(1)})}$ and $v \to 
u^{(1)}$ it is straightforward to end the
proof.
For $k>2$ we use the same argument as in the proof of theorem
\ref{theo:embedding}.
\finprfbis
\begin{theorem}
\label{embedding-co}
In the coset $\cA_{\fm|\fn}/\cJ_{k} \otimes \fD_{\fm|\fn}/\cI_{k}$,
 the coproduct takes the form
\beq
\Delta(D_{[1]}^{(k)}(u)) \equiv
T_{[2]}^{(k)}(u)\,D_{[1]}^{(k)}(u)\,(T_{[2]}^{-1})^{(k)}(\inv(u))
\mb{mod $\cJ_{k}$},
\eeq
where $[1]$ labels the space $\fD_{\fm|\fn}/\cI_{k}$, $[2]$ labels the 
space $\cA_{\fm|\fn}/\cJ_{k}$ and $\Delta$ is the coproduct of 
$\cA_{\fm|\fn}$. 
\end{theorem}
\prf
As in theorem \ref{theo:embedding}, we just do the proof for  $k=2$, the other
cases follow.
{From}
\ben
\Delta\Big(d_{ij}(u^{(1)})-\delta_{ij}\,\psi_1(u^{(1)})\,d_{11}(u^{(1)})\Big)
&=&
\sum_{a,b=1}^{\fm+\fn} \Big\{t_{ia}(u^{(1)})\,t'_{bj}(\inv(u^{(1)})) \otimes 
d_{ab}(u^{(1)})\nonu
&&-\delta_{ij}\,\psi_1(u^{(1)})\,t_{1a}(u^{(1)})\,t'_{b1}(\inv(u^{(1)})) 
\otimes d_{ab}(u^{(1)})\Big\}.
\nonumber
\een
and using the quotient $\cI_2$ and $\cJ_2$,
it follows:
\ben
&&\Delta\Big(d_{ij}(u^{(1)})-\delta_{ij}\,\psi_1(u^{(1)})\,d_{11}(u^{(1)})\Big)
=
\sum_{a,b=2}^{\fm+\fn} t_{ia}(u^{(1)})\,t'_{bj}(\inv(u^{(1)})) \otimes 
d_{ab}(u^{(1)})  +\nonu
&&\qquad+\, t_{i1}(u^{(1)})\,t'_{1j}(\inv(u^{(1)})) \otimes 
d_{11}(u^{(1)})-\delta_{ij}\,\psi_1(u^{(1)})\,t_{11}(u^{(1)})
\,t'_{11}(\inv(u^{(1)})) 
\otimes d_{11}(u^{(1)}).
\nonumber
\een
 Using the commutation relation (\ref{TRT}) for the second term we find:
\ben
\Delta\Big(d_{ij}(u^{(1)})-\delta_{ij}\,\psi_1(u^{(1)})\,d_{11}(u^{(1)})\Big) 
=\sum_{a,b=2}^{\fm+\fn}t_{ia}(u^{(1)})\,t'_{bj}(\inv(u^{(1)})) 
\otimes 
\Big(d_{ij}(u^{(1)})-\delta_{ij}\,\psi_1(u^{(1)})\,d_{11}(u^{(1)})\Big).
\nonumber
\een
Theorem \ref{theo:embedding} and lemma \ref{lem-1} allow us to 
generalise this result to each $k$.  
\finprfbis

\section{Highest weight representations\label{sec:hw}}

The fundamental point in using the ABA is to know a pseudo-vacuum 
for the model. In the mathematical framework 
it is equivalent to know a highest weight 
 representation for the algebra which underlies the model. 
Since the generators of the algebra $\fD_{\fm|\fn}$ can be constructed 
from the $\cA_{\fm|\fn}$ ones, see eq. (\ref{D-from-T}), 
we first describe how to construct highest repesentations for the infinite 
dimensional (graded) algebras
$\cA_{\fm|\fn}$ from highest  weight representation of the 
finite dimensional Lie subalgebras $gl(\fm|\fn)$ or 
$\cU_{q}(\fm|\fn)$.
Next, we show how these representations induce (for diagonal $K^-(u)$  
matrix) a representation for $\fD_{\fm|\fn}$ with same
highest weight vector.

\subsection{Finite dimensional representations of $\cA_{\fm|\fn}$} 

\begin{definition}
A representation  of $\cA_{\fm|\fn}$ is called \textit{highest 
weight} 
if there exists a nonzero vector $\Omega$ 
such that,
\begin{equation}
t_{ii} (u)\, \Omega = \lambda_{i}(u)\, \Omega \mb{and}
t_{ij} (u)\, \Omega = 0 \ \text{ for }\ i>j,   
\label{higvectU}
\end{equation}  
for some scalars $\lambda_{i}(u)$ $\in$ $\CC$ $[[u^{-1}]]$.
$\lambda(u)= (\lambda_{1}(u),\dots,\lambda_{\fm+\fn}(u))$ is 
called 
the highest weight and $\Omega$ the highest weight vector.
\end{definition}
It is known (see \cite{CP1,M1,Zhang,Ro}) that
any finite-dimensional irreducible representation of 
$\cA_{\fm|\fn}$ is  highest weight and that it contains a unique 
 (up to scalar multiples) highest weight vector.
To construct such representations, one uses the evaluation morphism, 
which relates the infinite dimensional algebra $\cA_{\fm|\fn}$ to its 
finite dimensional subalgebra $\cB_{\fm|\fn}$ (see \cite{BR1}). 
{From} the evaluation morphism $ev_{a}$ (with $a\in\CC$) and a highest weight 
representation 
$\pi_{\mu}$ of $\cB_{\fm|\fn}$ (where $\mu$ is a $\cB_{\fm|\fn}$ 
highest weight), one can construct a highest weight 
representation of $\cA_{\fm|\fn}$, 
called evaluation representation:
\begin{equation}
\rho^\mu_{a}=ev_{a}\,\circ\,\pi_{\mu}\ :\ \cA_{\fm|\fn}\ 
\stackrel{ev_{a}}{\longrightarrow}\ \cB_{\fm|\fn}\  
\stackrel{\pi_{\mu}}{\longrightarrow}\ \cV_{\lambda}\,.  
\end{equation}
The weight of this evaluation representation is given by
$\lambda(u)=\big(\lambda_{1}(u),\ldots,\lambda_{\fm+\fn}(u)\big)$, 
with
\begin{equation}
\lambda_{j}(u) =\begin{cases}
 u-a-(-1)^{[j]}\,\hbar\,\mu_j \mb{for} \cY(\fm|\fn) 
\\[1.2ex]
\displaystyle
(-1)^{[j]}\,\left(\frac{u}{a}\,\eta_{j}q^{\mu_{j}} - 
\frac{a}{u}\,\eta_{j}q^{-\mu_{j}}\right)
\mb{for}\wh\cU_q(\fm|\fn)
\end{cases}
\ j=1,\ldots,\fm+\fn,
\label{eq:Lambda-eval}
\end{equation}
where $\mu_{j}$, $j=1,\ldots,\fm+\fn$ are the weights of the 
$\cB_{\fm|\fn}$ 
representation.
More generally, one constructs tensor product of evaluation 
representations using the coproduct of $\cA_{\fm|\fn}$,
\ben
\Big(\otimes_{i=1}^{L}\,\rho^{\mu^{\langle 
i\rangle}}_{a_i}\Big)\, \circ 
\Delta^{(L)}\Big(T(u)\Big) = \rho^{\mu^{\langle 
1\rangle}}_{a_{1}}\Big(T(u)\Big)\dot{\otimes}\,
\rho^{\mu^{\langle 2\rangle}}_{a_{2}}\Big(T(u)\Big) 
\dot{\otimes}\cdots 
\dot{\otimes}\rho^{\mu^{\langle L\rangle}}_{a_{L}}\Big(T(u)\Big),
\label{eq:mono-repr}
\een
where $\mu^{\langle i\rangle}=(\mu^{\langle 
i\rangle}_{1},\ldots,
\mu^{\langle i\rangle}_{\fm+\fn})$, 
$i=1,\ldots,L$, are the weights of the $\cB_{\fm|\fn}$ 
representations. 
This provides a $\cA_{\fm|\fn}$ representation with weight,
\ben
\lambda_j(u)=\prod_{i=1}^{L}\lambda^{\langle i\rangle}_j(u)
\,,\qquad j=1,\ldots,\fm+\fn,
\label{VP}
\een
where $\lambda^{\langle i\rangle}_j(u)$ have the form 
(\ref{eq:Lambda-eval}).

%%%%%%%%%%%%%%%%%%%%%%%%%%%%%%%%%%%%%%%%%%%%

\subsection{Representations of $T^{-1}(u)$ from $T(u)$}

The construction of the finite dimensional representations for 
$T^{-1}(u)$ in relation with the $T(u)$ ones is different for the 
$gl(\fn)$ and the super symmetric case $gl(\fm|\fn)$.
For the  $gl(\fn)$ algebra, the representations are constructed 
using the quantum determinant
$qdet(T(u))$ and the comatrix
$\wh T(u)$ see \cite{M2}, while for the $gl(\fm|\fn)$ superalgebra, 
one uses the Liouville contraction, the quantum Berezinian 
$Ber(T(u))$ \cite{N} 
and crossing symmetry of $T(u)$.

We define for this section:
\ben
u_{\{k\}}&=&
\begin{cases}
u+\hbar k \\[1.2ex]
u\,q^{-k} 
\end{cases}\mb{and}
 f_{ij}(\sigma)\ =\
\begin{cases}
(-1)^{i+j+1}\,s(\sigma) &\mb{for $\cY(\fn)$ and 
$\cY(\fm|\fn)$}\\[1.2ex]
(-q)^{l(\sigma)+i-j}  &\mb{for $\wh\cU_q(\fn)$ and 
$\wh\cU_q(\fm|\fn)$}
\end{cases}
\een
where $s(\sigma)$ is the sign of the permutation $\sigma$ and $l(\sigma)$ its length.

\paragraph{$\cA_\fn$ case:}

We use the $\cA_\fn$ quantum determinant 
$qdet(T(u))$ which generates the center of $\cA_\fn$,
\beq
qdet(T(u))=\sum_{\sigma \in S_\fn} f_{00}(\sigma)\,
\prod_{i=1}^{\fn}t_{i \sigma(i)}(u_{\{i-\fn\}}), 
\label{qdet}
\eeq
and the quantum comatrix, 
\ben
\wh T(u)&=&\sum_{i j=1}E_{ij}\otimes\wh t_{ij}(u)\\
\wh t_{ij}(u)&=&\sum_{\sigma \in S_\fn} f_{ij}(\sigma)^{-1}\,t_{1 
a_{\sigma(1)}}(u_{\{2-\fn\}}) \dots t_{i-1 
a_{\sigma(i-1)}}(u_{\{i-\fn\}})\,t_{i+1 a_{\sigma(i)}}(u_{\{i+1-\fn\}}) \dots  t_{\fn 
a_{\sigma(\fn-1)}}(u) \nonumber \\
&&\mbox{with }\ (a_1,\dots,a_{\fn-1})=(1,\dots,j-1,j+1,\dots,\fn)
\een  
which obeys
$\wh T(u)\,T(u_{\{1-\fn\}})=qdet(T(u))$. 
This equation allows to relate $T^{-1}(u)$ to 
$\wh T(u)$ : 
\ben
T^{-1}(u) = \sum_{i,j=1}^{\fm+\fn}\,E_{ij} \otimes 
t'_{ij}(u)=\frac{\wh T(u_{\{\fn-1\}})}{qdet(T(u_{\{\fn-1\}}))}.
\een
To write the form of the highest weight irreducible 
representation for $T^{-1}(u)$, 
one first computes the action of  $qdet(T(u))$ and $\wh t_{ii}(u)$ on 
$\Omega$:
\ben
qdet(T(u))\,\Omega&=&\prod_{i=1}^{\fn}\lambda_i(u_{\{i-\fn\}})\,\Omega, \\
\wh t_{ii}(u)\,\Omega&=&\lambda_1(u_{\{2-\fn\}}) \dots 
\lambda_{i-1}(u_{\{i-\fn\}})\lambda_{i+1}(u_{\{i+1-\fn\}}) \dots 
\lambda_{\fn}(u)\,\Omega\,.
\een
Then, since $\wh t_{ij}(u)\,\Omega=0$ for $i>j$, one finds:
\ben
t'_{ii}(u)\,\Omega&=&\lambda'_i(u)\,\Omega \mb{with}
\lambda'_i(u)=\left( 
\prod_{k=1}^{i-1}\frac{\lambda_k(u_{\{k\}})}{\lambda_k(u_{\{k-1\}})} 
\right)\frac{1}{\lambda_i(u_{\{i-1\}})},
\nonu
t'_{ij}(u)\,\Omega\ &=&\ 0 \mb{if}  i>j .
\label{T-1-rep-n}
\een

\paragraph{$\cA_{\fm|\fn}$ case:}

First, one has to prove that $\Omega$ is a highest weight vector
of $T^{-1}(u)$. The proof is done in \cite{RS} 
for the super-Yangian case.
The quantum superalgebra $\wh\cU_q(\fm|\fn)$ case is done in the 
following theorem:
\begin{theorem}\label{theo:HWTinv}
For the quantum superalgebra $\wh\cU_q(\fm|\fn)$,
the highest weight vector $\Omega$ of $T(u)$ is also a highest weight vector of $T^{-1}(u)$. 
\ben
t'_{ii}(u)\,\Omega=\lambda'_i(u)\,\Omega \mb{and}
t'_{ij}(u)\,\Omega\ = 0 \mb{if}  i>j
\een
\end{theorem}
 \prf
To prove this theorem we must use the commutation relation between the modes 
of 
$T(u)=L^+(u)$ and 
$T^{-1}(u)=(L^+)^{-1}(u)=\sum_{i,j=1}^{\fm+\fn}\bar{L}^+_{ij}(u)\,.$
As $L^+(u)$ is a formal Taylor series in $u$, its inverse is also a formal Taylor series of $u$:
\ben
L^+_{ij}(u)= \sum_{n=0}^\infty L^{(n)}_{ij}\,u^{2n} \mb{and}
\bar{L}^+_{ij}(u)= \sum_{n=0}^\infty \bar{L}^{(n)}_{ij}\,u^{2n}.
\een
Projecting the commutation relation (\ref{TRT}) on the modes we find the following
relation:
\ben
 [ \bar{L}^{(p)}_{ij};L^{(q)}_{kl}\}&=&
\delta_{il} \sum_{a=1}^{\fm+\fn} (-1)^{([a]+[j])([a]+[k])}\,
\Big(c_{al}^+\,\sum_{b=0}^{p}L^{(q+b)}_{ka}\,\bar{L}^{(p-b)}_{aj}
-c_{al}^-\,\sum_{b=1}^{p}L^{(q+b)}_{ka}\,\bar{L}^{(p-b)}_{aj}\Big) \nonu
&-&\delta_{jk} \sum_{a=1}^{\fm+\fn} (-1)^{([i]+[l])([a]+[k])}\,
\Big(c_{ka}^+\,\sum_{b=0}^{p}\bar{L}^{(p-b)}_{ia}\,L^{(q+b)}_{al}
-c_{ka}^-\,\sum_{b=1}^{p}\bar{L}^{(p-b)}_{ia}\,L^{(q+b)}_{al}\Big), \nonu
\mb{with}&& 
 c_{al}^\pm=q^{\pm(1-2[l])}-q^{sign(l-a)(1-2[l])}.
\label{CRM}
\een
From the relation $T(u)\,T^{-1}(u)=T^{-1}(u)\,T(u)=\II$ we find:
\beq
\sum_{a=1}^{\fm+\fn}\sum_{q=0}^{p} (-1)^{([i]+[a])([a]+[j])}
\,\bar{L}^{(q)}_{ia}L^{(p-q)}_{aj}
=\sum_{a=1}^{\fm+\fn}\sum_{q=0}^{p} (-1)^{([i]+[a])([a]+[j])}
\,L^{(q)}_{ia}\bar{L}^{(p-q)}_{aj} 
 = \delta_{ij}\delta_{0p}.
\label{MI}
\eeq
We also know that $L^{(0)}_{ij}=0$ for $i<j$ and that $L^{(0)}_{ii}$ has an inverse.
First we prove the same properties for $\bar{L}^{(0)}$. From 
(\ref{MI}), we deduce:
\ben
\sum_{a=1}^{\fm+\fn}(-1)^{([i]+[a])([a]+[j])}\,L^{(0)}_{ia}\bar{L}^{(0)}_{aj}=0 
\mb{for} i\neq j\,.
\een
As $L^{(0)}_{ii} \neq 0$, we find for $i=1$, $\bar{L}^{(0)}_{1j}=0$. 
By induction on $i$,
we find  $\bar{L}^{(0)}_{ij}=0$ for $i<j$.
Then, 
\ben
\sum_{a=1}^{\fm+\fn}(-1)^{[i]+[a]}
\,L^{(0)}_{ia}\,\bar{L}^{(0)}_{ai}
=\sum_{a=1}^{\fm+\fn}(-1)^{[i]+[a]}
\,\bar{L}^{(0)}_{ia}L^{(0)}_{ai}=1 
\een
implies that $L^{(0)}_{ii}\,\bar{L}^{(0)}_{ii}=\bar{L}^{(0)}_{ii}\,L^{(0)}_{ii}=1$.

Now, we have to prove that $\Omega$ is a highest weight vector of
$\bar{L}_{ij}^{(p)}$. We already know
 that:
\ben
L_{ij}^{(p)}\,\Omega=0 \mb{for} i<j \mb{and} 
L_{ii}^{(p)}\,\Omega=\lambda_i^{(p)}\,\Omega\,.
\een 
We can write from (\ref{MI}) with $i>j$:
\ben
\sum_{q=0}^{p}\lambda_j^{(p-q)}\,\bar{L}^{(q)}_{ij}\,\Omega&=&-
\sum_{a=1}^{j-1}\sum_{q=0}^{p-1}(-1)^{([k]+[a])([i]+[j])}\,
\bar{L}^{(q)}_{ia}\,L^{(p-q)}_{aj}\,\Omega \,.
\een
To prove that $L_{ij}^{(p)}\,\Omega=0$ for $i<j$, 
we use a double induction, on $p$ and on $i$. 
We already proved 
 directly that $\bar{L}^{(0)}_{ij}\,\Omega=0$, $i<j$.
For $p=1$ we have:
\ben
\lambda_j^{(0)}\bar{L}^{(1)}_{ij}\,\Omega &=&-
\sum_{a=1}^{j-1}(-1)^{([k]+[a])([i]+[j])}
\,\bar{L}^{(0)}_{ia}\,L^{(1)}_{aj}\,\Omega 
=-
\sum_{a=1}^{j-1}(-1)^{([k]+[a])([i]+[j])}\,
[\bar{L}^{(0)}_{ia},L^{(1)}_{aj}\}\,\Omega. 
\nonumber
\een
Using the commutation relations (\ref{CRM}) we find:
\beq
[\bar{L}^{(0)}_{ia},L^{(1)}_{aj}\}\,\Omega = -q^{-1+2[a]}
\,(q-q^{-1})\,(-1)^{([i]+[j])[a]}
\sum_{b=1}^{a-1}(-1)^{[b]([i]+[b]+[j])}\,[\bar{L}^{(0)}_{ib},L^{(1)}_{bj}\}\,\Omega. 
\quad
\eeq
We get a triangular system in $a$, so that the property 
is proved for $p=1$ by induction on $a$. 
For a general $p$ we use the same method. 
% Let the property be true for $p-1$:  
%  \ben
% \sum_{q=0}^{p}\lambda_j^{(p-q)}\bar{L}^{(q)}_{ij}\,\Omega &=&-
% \sum_{a=1}^{j-1}\sum_{q=0}^{p-1}(-1)^{([k]+[a])([i]+[j])}
% \,[\bar{L}^{(q)}_{ia},L^{(p-q)}_{aj}\}\,\Omega. 
% \een
% Using the commutation relations (\ref{CRM}) we find:
% \ben
% [\bar{L}^{(q)}_{ia},L^{(p-q)}_{aj}\}\,\Omega &=&
% -q^{-1+2[a]}\,(q-q^{-1})\,(-1)^{([i]+[j])[a]+[i][j]}\sum_{b=1}^{a-1}
% \sum_{r=1}^{q}[\bar{L}^{(q-r)}_{ib},L^{(p-q+r)}_{bj}\}\,\Omega, \nonu
% \null [\bar{L}^{(0)}_{ia},L^{(q)}_{aj}\}\,\Omega &=&
% -q^{-1+2[a]}\,(q-q^{-1})\,(-1)^{([i]+[j])[a]+[i][j]}\,
% \sum_{b=1}^{a-1}[\bar{L}^{(0)}_{ib},L^{(q)}_{bj}\}\,\Omega. 
% \een
% Again, the system is triangular in $a$, hence the property is proved for each $p$.

Finally, we prove that $\bar{L}^{(p)}_{ii}\Omega=\bar{\lambda}_i^{(p)}\Omega$.
For $p=0$, from the equation $\bar{L}^{(0)}_{ii}L^{(0)}_{ii}=1$,
we already know that
 $\bar{L}^{(0)}_{ii}\Omega=(\lambda_i^{(0)})^{-1}\Omega$.
We prove the property of the general case by a double induction, assuming the property 
is true for $p-1$ and starting from:
\ben
\sum_{q=0}^{p}\lambda_i^{(p-q)}\bar{L}^{(q)}_{ii}\,\Omega &=&
\Big(\delta_{0,p}-
\sum_{a=1}^{i-1}\sum_{q=0}^{p-1}(-1)^{[i]+[a]}\,
[\bar{L}^{(q)}_{ia},L^{(p-q)}_{ai}\}\Big)\,\Omega. 
\een
From the commutation relation we get:
\ben
[\bar{L}^{(q)}_{ia},L^{(p-q)}_{ai}\}\,\Omega &=&
\sum_{b=1}^{a}\Big(c_{bi}^-\sum_{r=1}^{q}[\bar{L}^{(q-r)}_{ib},L^{(p-q+r)}_{bi}\}
-c_{ab}^+\sum_{r=0}^{q}[\bar{L}^{(q-r)}_{ib},L^{(p-q+r)}_{bi}\}\Big)\,\Omega, \nonu
\null [\bar{L}^{(0)}_{ia},L^{(q)}_{ai}\}\,\Omega &=&
-q^{-1+2[a]}\,(q-q^{-1})\,(-1)^{[a]}\,\sum_{b=1}^{a-1}
 [\bar{L}^{(0)}_{ib},L^{(q)}_{bi}\}\,\Omega=0.
\een
From the second equation, equals to zero by induction on $a$, we find:
\ben
[\bar{L}^{(q)}_{ia},L^{(p-q)}_{ai}\}\,\Omega &=&
-q^{-1+2[a]}\,(q-q^{-1})\,(-1)^{[a]}\,\sum_{b=1}^{a-1}
\sum_{r=1}^{q}[\bar{L}^{(q-r)}_{ib},L^{(p-q+r)}_{bi}\}\,\Omega.
\een
By induction on $q$ then on $a$ we find the last equality equals to zero.
Thus, we have the relation:
\ben
\sum_{q=0}^{p}\lambda_i^{(p-q)}\bar{L}^{(q)}_{ii}\,\Omega
&=&\delta_{0,p}\,\Omega
\een
It follows that $\bar{L}^{(q)}_{ii}\,\Omega=\Bar{\lambda}_i^{(q)}\,\Omega$ with 
$ \Bar{\lambda}_i^{(q)}$
rational function of  $ \lambda_i^{(q)},\dots,\lambda_i^{(0)}$.
 \finprf
 
Secondly, we use the crossing symmetry of the monodomy matrix 
$T_a(u)$ and 
the quantum Berezinian to give an explicit expression of the 
weight of $T^{-1}(u)$.
Let us introduce:
\ben
T^*(u)=(T^{-1}(u))^t.
\een
The crossing symmetry takes the form for $\cA_{\fm|\fn}$ 
(see \cite{RS,BR2}):
\ben
(T^t(u))^{-1}=\frac{1}{Z(u_{\{\fn-\fm\}})}\,M\,U\,
T^*(u_{\{\fn-\fm\}})\,U\,M^{-1}
\mb{with} U=\sum_{i=1}^{\fm+\fn} (-1)^{[i]}\,E_{ii}\,,
\een
where the Liouville contraction $Z(u)$ lies in the centre of 
$\cA_{\fm|\fn}$. It can be written in term of
the Berezinian, that itself relies on the quantum determinant (\ref{qdet}):
\ben
Ber(T(u))\! &=&\! qdet\Big(T^{(\fm)}(u_{\{\fm-\fn-1\}})\Big)
\, \wt{qdet}\Big((T^*)^{(\fn)}(u_{\{-\fn\}})\Big)\, =\, 
\prod_{i=1}^{\fm}\lambda_i(u_{\{i-\fn-1\}})
\prod_{i=1}^{\fn}\lambda'_{i+\fm}(u_{\{i-\fn-1\}})
\qquad\quad\\
Z(u) &=& \frac{Ber(T(u_{\{1\}}))}{Ber(T(u))}\mb{with}
\wt{qdet}(T^{(\fn)}(u))=\sum_{\sigma \in S_\fn} f_{00}(\sigma)^{-1}\,
\prod_{i=1}^{\fn}t_{i+\fm, \sigma(i)+\fm}(u_{\{\fn-i\}}), 
\een
with $T^{(k)}(u)=\II^{(k)}\,T(u)\,\II^{(k)}$, $\forall\,k$ 
and $\II^{(k)}$ defined in (\ref{eq:Ik}). 
In $\cA_{0|\fn}\subset\cA_{\fm|\fn}$, 
we have: 
\ben
(T^{(\fn)}(u)^t)^{-1}=z(u)\,M\,T^{(\fn)*}(u_{\{\fn\}})\,M^{-1},
\een
where $z(u)$, the $\cA_{0|\fn}$ Liouville contraction, 
 can be writen in term of the quantum determinant:
\ben
z(u)&=&
% \left.
\frac{\wt{qdet}(T^{(\fn)}(u_{\{1\}}))}{\wt{qdet}(T^{(\fn)}(u))}
% \,\right \vert_{q \to q^{-1}}
\ =\ 
\prod_{i=1}^{\fn} \frac{\lambda_{i+\fm}(u_{\{\fn-i+1\}})}{\lambda_{i+\fm}(u_{\{\fn-i\}})}.
\een
\begin{lemma} For the superalgebra $\cA_{\fm|\fn}$, we have: 
\ben
t'_{ii}(u)\,\Omega&=&\lambda'_i(u)\,\Omega \mb{and}
t'_{ij}(u)\,\Omega\ =\ 0 \mb{for}  i>j ,
\nonumber\\[1.2ex]
\mb{with} \lambda'_i(u) &=&
\begin{cases}
\displaystyle
\frac{1}{\lambda_i(u_{\{i-1\}})}\,
\left(\prod_{k=1}^{i-1}\frac{\lambda_k(u_{\{k\}})}{\lambda_k(u_{\{k-1\}})} 
\right) \mb{for}  1\leq i\leq \fm 
\\[2.em]
\displaystyle
\frac{Z(u)}{\lambda_{i}(u_{\{2\,\fm-i\}})}\,
\left(\prod_{k=i+1}^{\fm+\fn}
\frac{\lambda_{k}(u_{\{2\,\fm+1-k\}})}{\lambda_k(u_{\{2\,\fm-k\}})} 
\right) 
\mb{for}  \fm+1\leq i\leq \fm +\fn
\end{cases}.
\een
\end{lemma}
\prf
From  theorem \ref{theo:HWTinv}, we have, 
\ben
T^{-1}(u)\,\Omega&=&\left (
\begin{array}{cc} T^{-1}(u)^{(\fm)} & *
\\  0&  T^{-1}(u)^{(\fn)} 
\end{array} \right )\Omega,
\label{eq:hwTinv}
\\
T^*(u)\,\Omega&=&\left(
\begin{array}{cc} T^*(u)^{(\fm)} & 0
\\  * &  T^*(u)^{(\fn)} 
\end{array} \right )\Omega\,.
\label{eq:hwT*}
\een
Multiplying (\ref{eq:hwTinv}) by $T(u)$ and (\ref{eq:hwT*}) by 
$UM^{-1}\,T^{t}(u_{\{\fn-\fm\}})\,MU$, one gets:
$$
T^{(\fm)}(u)\,(T^{-1})^{(\fm)}(u)\,\Omega = \Omega \mb{and}
(M^{-1})^{(\fn)}\,(T^t)^{(\fn)}(u_{\{\fn-\fm\}})\,M^{(\fn)}\,(T^*)^{(\fn)}(u)\, \Omega \ =\ Z(u)\,\Omega.
$$
Finally, upon multiplication by $T^{(\fm)}(u)^{-1}$ and  
$(T^t)^{(\fn)}(u)^{-1}$, one is led to
\beq
(T^{-1})^{(\fm)}(u)\,\Omega = \big(T^{(\fm)}\big)^{-1}(u)\,\Omega \mb{and}
(T^*)^{(\fn)}(u)\, \Omega \ =\ 
{Z(u)}\,{z(u_{\{\fm-\fn\}})}\,\big(T^{(\fn)}\big)^{*}(u_{\{\fm\}})\,\Omega
\eeq
that gives the lemma.
\finprfbis

\subsection{Finite dimensional representations of $\fD_{\fm|\fn}$ 
from $\cA_{\fm|\fn}$ ones} 

For the study of the representations of the reflection algebra, we
follow essentially the lines given in \cite{MoRa} for the
reflection algebra based on the Yangian of $gl(\fn)$ and in 
\cite{RS} for the
reflection algebra based on the super-Yangian of $gl(\fm|\fn)$.

\begin{theorem}\label{theo:valB}
If $\Omega$ is a highest weight vector of $\cA_{\fm|\fn}$, with 
eigenvalue
$( \lambda_{1}(u),\ldots, \lambda_{\fm+\fn}(u))$, then, when 
$K^-(z)=diag\big(\kappa_1(u),\ldots,\kappa_{\fm+\fn}(u)\big)$, 
$\Omega$ is also a highest weight 
vector for $\fD_{\fm|\fn}$,
\ben
d_{ij}(u)\,\Omega =0  \mb{for}  i>j,
\mb{and}
d_{ii}(u)\,\Omega =\Lambda_i(u)\,\Omega, 
\een
with eigenvalues:
\ben
\Lambda_i(u)&=&\cK_{i}(u)\,\lambda_{i}(u)\,\lambda'_{i}(\inv(u))
+\sum_{k=1}^{i-1}
\psi_{k}(u^{\wb{(1 \dots k-1)}})\,
\cK_{k}(u)\,\lambda_{k}(u)\,\lambda'_{k}(\inv(u)),
\label{eq:valprD}\\
\cK_{i}(u)&=&\kappa_{i}(u)-\sum_{k=1}^{i-1}\kappa_{k}(u)
\frac{\fw_{ik}(u^{\wb{(1 \dots k-1)}},\inv(u^{\wb{(1 \dots k-1)}}))}
{\fa_{i-1}(u^{\wb{(1 \dots i-2)}},\inv(u^{\wb{(1 \dots 
i-2)}}))}\,q^{i-k-1-2\sum_{l=k+1}^{i-1}[l]}.
\label{eq:grKapp}
\end{eqnarray}
\end{theorem}
\prf
 First, we prove $d_{ij}(u)\,\Omega=0$ for
$j < i$.
One computes:
\begin{eqnarray}
d_{ij}(u)\,\Omega &=&
\sum_{a=1}^{j-1}\,(-1)^{([i]+[a])([a]+[j])}\,
\kappa_{a}(u)\,t_{ia}(u)\,t'_{aj}(\inv(u))\,
\Omega
\label{Bom}
\\ 
&=&-\sum_{a=1}^{j-1} 
\kappa_{a}(u)\,\Big[t'_{aj}(\inv(u))\,,\,t_{ia}(u)\Big\}
\,\Omega. \nonumber
\end{eqnarray}
Applying the super-commutator on $\Omega$ with the constraint 
$a \leq j < i$, one gets:
\begin{equation}
\Big[t'_{aj}(u)\,,\,t_{ia}(\inv(u))\Big\}\,\Omega =
-\sum_{b=1}^{j-1}\frac{\fw_{ba}(u,\inv(u))}{\fb(u,\inv(u))}\,
\Big[t'_{bj}(\inv(u))\,,\,t_{ib}(u)\Big\}\,\Omega.
\end{equation}
Considering the case $a=j$, one obtains:
\begin{equation}
\sum_{b=1}^{j-1}\,\Big[t'_{bj}(\inv(u))\,,\,t_{ib}(u)\Big\}\,\Omega
=0\,.
\end{equation}
Plugging this result in the former equation, we get :
\begin{equation}
\Big[t'_{aj}(u)\,,\,t_{ia}(\inv(u))\Big\}\,\Omega =
\frac{\fw_{a-1,a}(u,\inv(u))-\fw_{a+1,a}(u,\inv(u))}{\fa_a(u,\inv(u))
-\fw_{a-1,a}(u,\inv(u))}
\sum_{b=a+1}^{j-1}\,\Big[t'_{bj}(\inv(u))\,,\,t_{ib}(u)\Big\}\,\Omega.
\end{equation}
By iteration ($a=j-1$,..., $a=1$) one finds:
\ben
\Big[t'_{aj}(u)\,,\,t_{ia}(\inv(u))\Big\}\,\Omega=0,
\een
which proves that
$d_{ij}(u)\,\omega=0\,,\ j<i\,$.
\\
\\
Secondly, we prove $d_{ii}(u)\,\Omega=\,\Lambda_i(u)\,\Omega$.
Acting on $\Omega$ with $d_{ii}(u)$ one obtains:
\ben
d_{ii}(u)\,\Omega 
&=&\kappa_{i}(u)\lambda_i(u)\lambda'_i(\inv(u))\,\Omega
+\sum_{a=1}^{i-1}\,(-1)^{[i]+[a]}\,
\kappa_{a}(u)\,t_{ia}(u)\,t'_{ai}(\inv(u))\,
\Omega.
\label{d-v}
\een
One can restrict this problem to the computation of 
$t_{ia}(u)\,t'_{ai}(\inv(u))\,\Omega$
for $i>a$ in term of the eigenvalues 
$\lambda_i(u)\lambda'_i(\inv(u))$. From the relation (\ref{TRT}), we get
\ben
(-1)^{[i]+[a]}\,t_{ia}(u)\,t'_{ai}(\inv(u))\,\Omega&=&
 -\sum_{b=1}^{i}\, (-1)^{[i]+[b]}\,
 \frac{\fw_{ba}(u,\inv(u))}{\fb(u,\inv(u))}\, 
t_{ib}(u)\,t'_{bi}(\inv(u)) 
\,\Omega
\nonumber \\
&&+ \sum_{b=1}^{a} \frac{\fw_{ib}(u,\inv(u))}{\fb(u,\inv(u))}\, 
t'_{ab}(\inv(u))\,t_{ba}(u) \, \Omega.
\een
Applying (\ref{TRT}) on $\Omega$ for $i=j=k=l$, 
one finds the identity:
\ben
\sum_{b=1}^{i-1} \frac{\fw_{ib}(u,\inv(u))}{\fb(u,\inv(u))}\, 
t'_{ib}(\inv(u))
\,t_{bi}(u) \, \Omega=\sum_{b=1}^{i-1} \, (-1)^{[i]+[b]}\,
\frac{\fw_{bi}(u,\inv(u))}{\fb(u,\inv(u))}\, 
t_{ib}(u)\,t'_{bi}(\inv(u)) 
\,\Omega \,.\nonumber
\een
Let $F_{ij}=t_{ij}(u)\,t'_{ji}(\inv(u))\Omega$. Using the two 
previous equations, one finds for $j<i$:
\ben
F_{ij} &=& (-1)^{[i]+[j]}\,
\frac{\fw_{ij}(u,\inv(u))}{\fb(u,\inv(u))}\,(F_{jj}-F_{ii})
+\sum_{a=1}^{j-1}\,(-1)^{[i]+[a]}\,
\frac{\fw_{ai}(u,\inv(u))}{\fb(u,\inv(u))}\,F_{ja}
\nonu
&&-\sum_{a=1}^{i-1}\,(-1)^{[j]+[a]}\,
\frac{\fw_{aj}(u,\inv(u))}{\fb(u,\inv(u))}\,F_{ia}.
\een
It is then easy (but lengthy) to show that the solution is:
\begin{eqnarray}
 F_{ij} &=& (-1)^{[i]+[j]}\fm_{ij}(u^{\wb{(1\dots 
j-1)}},\inv(u^{\wb{(1\dots j-1)}}))
 \Big[\frac{F_{jj}}{\fa_{j}(u^{\wb{(1\dots j-1)}},
\inv(u^{\wb{(1\dots j-1)}}))} 
\nonu
&&-
\frac{q^{(i-j-1-2\sum_{a=j+1}^{i-1}[a])}}
{\fa_{i-1}(u^{\wb{(1\dots i-2)}},\inv(u^{\wb{(1\dots i-2)}}))}
\,F_{ii}\,
\nonu 
 &&-\sum_{a=j+1}^{i-1} \frac{\fm_{ia}(u^{\wb{(1\dots 
a-1)}},\inv(u^{\wb{(1\dots a-1)}}))
\,q^{(a-j-1-2\sum_{l=j+1}^{a-1}[l])}}
{\fa_{j}(u^{\wb{(1\dots a-1)}},\inv(u^{\wb{(1\dots 
a-1)}}))\,\fa_{a-1}(u^{\wb{(1\dots a-2)}},\inv(u^{\wb{(1\dots 
a-2)}}))}\,F_{aa}\Big].
 \end{eqnarray}
One must use relations (\ref{eq:fct-id1})-(\ref{eq:fct-id2}) between functions.
Plugging the value of $F_{ik}$ into the equation (\ref{d-v}), after 
some
rearrangement one gets the eigenvalues $\Lambda_i(u)$.
\finprfbis

\section{Algebraic Bethe ansatz for $\fD_{\fm|\fn}$ with $\fm+\fn=2$ 
\label{sec:ABA}}
In this section, we remind the framework of the Algebraic Bethe 
Ansatz (ABA) \cite{STF}
introduced  in order to compute transfer matrix
eigenvalues and eigenvectors.  For $\fm+\fn=2$, one can 
consider three different 
algebras: $\fD_{0|2},\fD_{2|0}$ and $\fD_{1|1}$.
The method follows the same steps as the closed chain case, up to a 
preliminary step. 
We write the monodromy matrix in the following
matricial form:
\beq
D(u) = \left (
\begin{array}{ccc} d_{11}(u) && d_{12}(u)
\\  d_{21}(u) && d_{22}(u)
\end{array} \right ).
\eeq
In the open case the transfer matrix have the form: 
\ben
d(u) &=& str\big(K^+_a(u)\,D_a(u)\big) =
(-1)^{[1]}\,m_1\,k(u)\,d_{11}(u)  + 
(-1)^{[2]}\,m_2\,d_{22}(u), \\
K^+(u) &=& M\,K(u).
\een
The matrix $K$ is construct 
from the solution (\ref{eq:Kdiag}):   
\beq
K(u)=\begin{cases}
\II &\mbox{ for } a_{+}=0 \\
\mbox{diag}\,(k(u),1) &\mbox{ for } a_{+}=1
\end{cases}
\ \mbox{ with }\
 k(u)=\begin{cases}
\displaystyle
\frac{-u-\frac{\fm-\fn}{2}\hbar-c_+}{u+\frac{\fm-\fn}{2}\hbar-c_+}
\ \mbox{ for }\cY(\fm|\fn)\\[2.4ex]
\displaystyle
\frac{u^{-2}\,q^{-\frac{\fm-\fn}{2}}-c_+^2}{u^2\,q^{\frac{\fm-\fn}{2}}-c_+^2}
\ \mbox{ for }\wh\cU_q(\fm|\fn)
\end{cases}.\
\label{eq:k-naba}
\eeq
Remark that for the particular case $\fm+\fn=2$, the form chosen for 
$K^{+}(u)$ exhausts all possible diagonal solutions. We recall that 
for $K^-(u)$ we keep the general diagonal solution (\ref{eq:Kdiag}).
Let $\Omega$ be the pseudo-vacuum state:
\ben
d_{11}(u) \, \Omega  &=& \Lambda_1(u) \,\Omega \mb{,}
d_{22}(u) \, \Omega  \,=\, \Lambda_2(u) \,\Omega \mb{,}
d_{21}(u) \, \Omega  \,=\,  0 \,.
\een
Looking at the commutation relations (\ref{RCP}) for $\fm+\fn=2$, one 
can see that the $d_{22}(u)\,d_{12}(v)$ exchange relation is not symmetric 
to the $d_{11}(u)\,d_{12}(v)$ one. In order to compensate this 
asymmetry, we perform a change of basis and a shift, 
\ben
d_{11}(u^{(1)}) &=& \widehat d_{11}(u)\mb{,} d_{12}(u^{(1)})=\widehat 
d_{12}(u)
\mb{,}
d_{21}(u^{(1)})=\widehat d_{21}(u),\\
d_{22}(u^{(1)}) &=& \widehat d_{22}(u)+\psi_1(u^{(1)})\,\widehat d_{11}(u).
\een
The function $\psi(u)$ is chosen in such a way that it leads to symmetric exchange relations:
\ben
 \widehat d_{12}(u)\,\widehat d_{12}(v)&=& 
\begin{cases}  
 \widehat d_{12}(v)\,\widehat d_{12}(u), \mbox{ for 
$\fD_{2|0},\fD_{0|2}$} \\[1.2ex]
 \fh(u,v)\,\widehat d_{12}(v)\,\widehat d_{12}(u), \mbox{ for 
$\fD_{1|1}$} 
\end{cases} ,
\label{eq:d12d12hat}\\
\widehat d_{11}(u)\, \widehat d_{12}(v)&=&
\ff_1(u,v)\,\widehat d_{12}(v)\, \widehat d_{11}(u)
+\fg_1(u,v)\,\widehat d_{12}(u)\, \widehat d_{11}(v)
+\fh_1(u,v)\,\widehat d_{12}(u)\, \widehat d_{22}(v),
\label{ExchABF} \quad
\\
\widehat d_{22}(u)\, \widehat d_{12}(v)&=& \wt \ff_2(u,v)\,
 \widehat d_{12}(v) \, \widehat d_{22}(u)
+\wt \fg_2(u,v)\,\widehat d_{12}(u)\, \widehat d_{22}(v) 
 +\wt \fh_2(u,v)\, \widehat d_{12}(v)\, \widehat d_{11}(u).
\label{ExchABFbis} 
\een
The explicit form of the functions appearing above is given in 
appendix \ref{app:fct}.
In the new basis, $\Omega$ is still a pseudo-vacuum:
\ben
\widehat d_{11}(u) \, \Omega  &=&\wh{\Lambda}_1(u)\,\Omega= 
\Lambda_1(u^{(1)}) \,\Omega \mb{,}
\widehat d_{21}(u) \, \Omega  \,=\,  0,\\
\widehat d_{22}(u) \, \Omega  &=& \widehat \Lambda_2(u) 
\,\Omega \,=\,\big(\Lambda_{2}(u^{(1)})-\psi_1(u^{(1)})\,
\Lambda_1(u^{(1)}) \big)\,\Omega
 \,.
\een
and we can use the algebraic Bethe ansatz as in the closed chain 
case. 
The transfer matrix rewrites:
\beq
d(u^{(1)}) =
\Big((-1)^{[1]}\,m_1\,k(u^{(1)})+(-1)^{[2]}\,m_2
\,\psi_1(u^{(1)})\Big)\,
\widehat d_{11}(u)+
(-1)^{[2]}\,m_2\,\widehat d_{22}(u)
\equiv \widehat d(u)
\eeq
Applying $M$ creation operators $\wh d_{12}(u_{j})$ on the pseudo vacuum we 
generate a Bethe vector:
\beq
\Phi(\{u\}) \, = \widehat d_{12} (u_1) \dots  \widehat d_{12} (u_M) 
\,\Omega.
\eeq
Demanding $\Phi(\{u\})$ to be an eigenvector of $\widehat d(u)$ leads
to a set of algebraic relations on the parameters 
$u_1,\dots,u_M$, the so-called Bethe equations.
The relation (\ref{eq:d12d12hat}) between creation operators proves the invariance (up to 
a function for $\fD_{1|1}$) of the
Bethe vector under the reordering of creation operators. 
This
condition is usefull to compute the unwanted terms from the 
action of $\widehat d(u)$ on  $\Phi(\{u\})$. 
We compute the action of $\widehat d_{11}(u)$ on $\Phi(\{u\})$,
\ben
\widehat d_{11} (u)\,\Phi(\{u\}) &=&  \prod_{k=1}^{M} \ff_1(u,u_k)
\,\wh{\Lambda}_1(u)\,\Phi(\{u\})\nonu
&+& \sum_{k=1}^{M}\Big(  
M_k(u,\{u\})\,\wh{\Lambda}_1(u_k) + N_k(u,\{u\})\,\wh{\Lambda}_2(u_k) 
\Big)\,\Phi_k(u,\{u\}),
 \nonumber \\
\Phi_k(u,\{u\})&=&
\widehat d_{12}(u_1) \dots\widehat d_{12}(u_k \to u) 
\dots
 \widehat d_{12} (u_M)\,\Omega.
\label{ABBO}
\een
 where the notation $\widehat d_{12}(u_k \to u)$ is used to indicate 
the position of $\widehat d_{12}(u)$ in the ordered product.
 The form of $M_1(u;\{u\})$ and $N_1(u;\{u\})$  is easily computed.
The other polynomials $M_k(u;\{u\})$ and $N_k(u;\{u\})$ are
 then computed using the 
commutation relation between the $\widehat d_{12}(u)$ operators 
 and puting $\widehat d_{12}(u_{k})$ on the left.
We get:
$$M_k(u,\{u\})=\fg_1(u,u_k)\prod_{i\neq k}^{M}\ff_1(u_k,u_i) \mb{and}
N_k(u,\{u\})=\fh_1(u,u_k)\prod_{i\neq k}^{M} \wt \ff_2(u_k,u_i).$$
Similarly, we compute the action of $\widehat d_{22}(u)$ on 
$\Phi(\{u\})$,
\ben
\widehat d_{22} (u)\,\Phi(\{u\}) &=& \left(
\prod_{k=1}^{M} \wt \ff_2(u,u_k)\right)
\,\widehat \Lambda_2(u)\,\Phi(\{u\})+ 
\label{DBBO}\\
&&+ \sum_{k=1}^{M} 
\Big(O_k(u,\{u\})\,\widehat \Lambda_2(u_k)+ 
P_k(u,\{u\})\,\wh\Lambda_1(u_k) \Big)\,
\Phi_k(u,\{u\}),
\nonu
O_k(u,\{u\}) &=&\wt \fg_2(u,u_k)\prod_{i\neq k}^{M} \wt 
\ff_1(u_k,u_i) \mb{and}
P_k(u,\{u\})= \wt \fh_2(u,u_k)\prod_{i\neq k}^{M}\ff_1(u_k,u_i).
\quad
\een
Demanding $\Phi(\{u\})$ to be an eigenvector of $\widehat d(u)$ 
corresponds to the cancelling of the so-called `unwanted terms'
carried by the vectors
 $\Phi_k(u,\{u\})$.
In this way, we get
the Bethe equations:
\ben
\frac{\wh \Lambda_1(u_k)}{\wh \Lambda_2(u_k)} &=&
\chi_1(u_k)\prod_{i\neq k}^{M}\frac{\wt
\ff_2(u_k,u_i)}{\ff_1(u_k,u_i)}\,,\quad k=1,\ldots,M\,.
\een
Remark that the r.h.s. depends only on the structure constants of the 
(super)algebra under consideration, while the l.h.s. encodes the 
representations entering the spin chain.
Then, the eigenvalues of the transfer 
matrix read:  
\ben
 \widehat d(u)\,\Phi(\{u\}) &=& 
\wh{\Lambda}(u;\{u\})\, \Phi(\{u\}), \\
\wh{\Lambda}(u;\{u\}) &=& 
\Big((-1)^{[1]}\,m_{1}\,k(u^{(1)})
+(-1)^{[2]}\,m_{2}\,\psi_1(u^{(1)})\Big)\,
\wh{\Lambda}_1(u)\,\prod_{k=1}^{M} 
\ff_1(u,u_{k}) \nonu
&&+\,(-1)^{[2]} \,m_2\,\widehat \Lambda_2(u)\,
 \prod_{k=1}^{M} \wt \ff_2(u,u_k)\,.
\een
Note that Bethe equations correspond to the vanishing
of the residue of  $\Lambda(u;\{u\})$. 
This is the tool used in 
analytical Bethe ansatz \cite{reshe} to obtain Bethe equations, 
see e.g. \cite{ACDFR,RS}.
%%%%%%%%%%%%%%%%%%%%%%%%%%%%%%%%%%%% 
\section{Nested Bethe ansatz\label{sec:NBA}}
%%%%%%%%%%%%%%%%%%%%%%%%%%%%%%%%%%%%
\subsection{Preliminaries}

The method, called the Nested Bethe Ansatz (NBA), consists in a
recurrent application of the ABA to express higher 
rank solutions using the lower ones. It has been introduced in 
\cite{KR} for the periodic case. The same method can be used for the
boundary case.
In this way, we can compute the eigenvalues, eigenvectors and Bethe
equations of the $\fD_{\fm|\fn}$ model from the ones of 
$\fD_2$ or $\fD_{1|1}$ model. 
Although we are in a 
(tensor product of) representation(s) of 
$\fD_{\fm|\fn}$, we will loosely keep writing $d_{ij}(u)$ the 
representation of the operators $d_{ij}(u)$, assuming that the 
reader will understand that when $d_{ij}(u)$ applies to the highest 
weight 
$\Omega$, it is in fact its (matricial) representation that is used.
Another way to understand this method in an algebraic way is to
work in the coset of the $\fD_{\fm|\fn}$ algebra by the left ideal 
$\cI_{\fm+\fn}$.

We consider now the open case with general diagonal boundary condition 
(\ref{eq:Kdiag}) for $K^-(u)$, and $K^+(u)$ of the form: 
\ben
K^+(u)&=&M\,K(u)\,,\mb{with}
K(u)\,=\,\begin{cases}
\II &\mb{for} a_{+}=0 \\
\mbox{diag}\,\Big(k(u),1,\dots,1\Big) &\mb{for} a_{+}=1
\end{cases}\,,\nonumber
\een
where the 
function $k(u)$ is defined in (\ref{eq:k-naba}).
The matrix $K^+(u)=M\,K(u)$ is the only solution we can use to perform 
the NBA up to the end (see remarks \ref{rmk:NABA} and \ref{rmk:NABA2} below).
We decompose the monodromy matrix in the following 
form  (in the $End(\mathbb{C}^{\fm+\fn})$ auxiliary space),
 \beq
D(u) =
 \left ( \begin{array}{cc} 
d_{11}(u) & B^{(1)}(u)  \\ 
C^{(1)}(u)  & D^{(2)}(u) 
\end{array} \right),
\eeq
where $B^{(1)}(u)$ (resp. $C^{(1)}(u)$) is a row (resp. column) 
vector 
of $\CC^{\fm+\fn-1}$, and $D^{(2)}(u)$ is a matrix of 
$End(\CC^{\fm+\fn-1})$.

Then, $D^{(2)}(u)$ is itself decomposed in the same way, and more 
generally, for a given $k$ in $\{1,\dots,\fm+\fn-2\}$,
we gather  the generators
$d_{kj}(u)$, (resp. $d_{jk}(u)$) $j=k+1,\ldots,\fn+\fm$,  in a row 
(resp. column) vector of 
$\CC^{\fm+\fn-k}$
and $d_{ij}(u)$, $i,j\geq k$, into a matrix of $End(\CC^{\fm+\fn-k})$:

\begin{eqnarray}
B^{(k)}(u) &=& \sum_{j=k+1}^{\fm+\fn}e^t_{j} \otimes d_{kj}(u)
\mb{and} 
C^{(k)}(u) = \sum_{j=k+1}^{\fm+\fn}e_{j} \otimes d_{jk}(u),
\\
D^{(k+1)} (u) &=& \sum_{i,j=k+1}^{\fm+\fn}\,E_{ij} \otimes d_{ij}(u),
\\[2.1ex]
D^{(k)}(u) &=& \left ( \begin{array}{cc} 
d_{kk}(u) & B^{(k)}(u)  \\ 
C^{(k)}(u)  & D^{(k+1)}(u) 
\end{array} \right).
\label{eq:mono-k}
\end{eqnarray}
We decompose the transfer matrix in the same way:
\ben
d(u) &=& d^{(1)}(u) = (-1)^{[1]}\,m_1\,k(u)\,d_{11}(u)+d^{(2)}(u)\,,\nonu
d^{(k)}(u) &=& str\Big(M^{(k)}_a\,D^{(k)}(u)\Big) = 
(-1)^{[k]}\,m_k\,d_{kk}(u)+d^{(k+1)}(u),\\
M^{(k)} &=& \II^{(k)}\,M\,\II^{(k)}\,.
\een
At each step of the recursion, 
we make a transformation of the operator and a shift of the
spectral parameter:
\ben
d_{kk}(u^{(k)})&=&\wh d_{kk}(u)\,, \qquad
B^{(k)}_a(u^{(k)})\,=\,\wh B^{(k)}_a(u), \nonu
D^{(k+1)}_a(u^{(k)})&=&\wh D^{(k+1)}_a(u)+
\psi_k(u^{(k)}) \,\II^{(k+1)}_a  \otimes \wh d_{kk}(u). 
\een
The commutation relations  
for these operators remain similar for each $k$:
  \ben
&&\wh B^{(k)}_a(u)\,\wh B^{(k)}_b(v)= 
(-1)^{[k]}\frac{\fa_{k+1}(u^{(k)},v^{(k)})}{\fa_{k}(u^{(k)},v^{(k)})}
\wh B^{(k)}_b(v)\, \wh B^{(k)}_a(u)\,\RR_{ba}^{(k+1)}(u,v), 
\label{eq:comBB}\\
&&\wh d_{kk}(u)\, \wh B^{(k)}_b(v)=
\ff_k(u,v) 
\wh B_b(v) \wh d_{kk}(u) + \fg_k(u,v)\wh B^{(k)}_b(u) 
\,\wh d_{kk}(v) +
\nonu
&&\qquad\qquad +\frac{\fh_k(u,v)}{\fe_{k+1}(v)}\,\wh B^{(k)}_b(u)\,
str_a\Big(M_a^{(k+1)}\,\wb{\RR}^{(k+1)}_{ab}(v,v)\,
\wh D^{(k+1)}_a(v)\,\RR^{(k+1)}_{ba}(v,v)\Big), 
\label{eq:comdkkB} \\
&& str_a\Big(M_a^{(k+1)}\,\wh D^{(k+1)}_a(u)\Big)\, \wh B^{(k)}_b(v)=
\wt \fh_{k+1}(u,v)\,\fe_{k+1}(u)\,\wh 
B^{(k)}_b(u)\,\wh d_{kk}(v)  \nonumber \\
&&\qquad\qquad +\frac{\wt \fg_{k+1}(u,v)
\,\fe_{k+1}(u)}{\fe_{k+1}(v)}\,\wh B^{(k)}_b(u)
\,str_a\left(M_a^{(k+1)}\,\wb{\RR}^{(k+1)}_{ab}(v,v)
\wh D^{(k+1)}_a(v)\RR^{(k+1)}_{ba}(v,v)\right)+ \nonu
 &&\qquad\qquad +\wt \ff_{k+1}(u,v)\,\wh B^{(k)}_b(v) 
 \,str_a\left(M_a^{(k+1)}\,\wb{\RR}^{(k+1)}_{ab}(u,v)
 \wh D^{(k+1)}_a(u) \RR^{(k+1)}_{ba}(u,v)\right).
\label{eq:comDB}
\een

\begin{rmk}\label{rmk:NABA}
The commutation relations (\ref{eq:comdkkB}) and (\ref{eq:comDB}) impose
 the restriction on the $K^+(u)$ matrix. 
The direct use of the reflection 
equation leads to a matrix $\wb{\RR}^{(k+1)}_{ab}(u,u)$ in 
(\ref{eq:comdkkB}) and (\ref{eq:comDB}).
The change  $\wb{\RR}^{(k+1)}_{ab}(u,u)\,\to\,\wb{\RR}^{(k+1)}_{ab}(v,v)$ 
in the commutation relation is allowed by
equality (\ref{sol}) which shows that the dependence in $u$ is a 
scalar function. If the $K^+(u)$ matrix is not from a \textbf{NABA 
couple}, equation (\ref{sol}) cannot be used to get (\ref{eq:comdkkB}) 
and (\ref{eq:comDB}) in their present form.
Without this form, the nesting cannot be performed (see also 
remark \ref{rmk:NABA2}).  
\end{rmk}

At each step $k=1,\ldots,\fm+\fn-1$ of the nesting, we will introduce 
a family of Bethe 
parameters 
$u_{kj}$, $j=1,\ldots,M_{k}$, 
the number $M_{k}$ of these parameters being a free integer. The 
partial unions of these families will be noted as,
\begin{equation}
\{u_{\ell}\}=\bigcup_{i=1}^{\ell}\,
\{u_{ij}\,,\ j=1,\ldots,M_{i}\},
\end{equation}
so that the whole family of Bethe parameters is
$\{u\}=\{u_{\fm+\fn-1}\}$.

\subsection{First step of the construction \label{first}}

{From} the definition of the highest weight, $C^{(1)}(u)$ annihilates 
the pseudo-vacuum $\Omega$
and we can use $B^{(1)}(u)$ as a creation operator. However, since 
$B^{(1)}(u)$ contains only $d_{1j}(u)$ operators, it is clear that we 
need to act on several vectors to describe the whole 
representation with highest weight $\Omega$. The NBA 
spirit is to 
construct these different vectors as Bethe vectors of an
$\fD_{\fm-1|\fn}$ chain that is related to the chain we start with.

More generally, at 
each step $k$ corresponding to the decomposition (\ref{eq:mono-k})
of the monodromy 
matrix and to the transformation of the operator of the corresponding 
algebra $\fD_{\fm-k|\fn}$, 
we use (a suitable refinement of) $B^{(k)}(u)$ 
as a creation operator acting on a 
 set of (to be defined) vectors. These vectors are constructed as 
 Bethe vectors of a $\fD_{\fm-k-1|\fn}$ chain.

At the first step of the recursion, the Bethe vectors have the 
form  
\ben
\Phi(\{u^{(1)}\}) &=& \wh B^{(1)}_{a^1_1}(u_{11}) \dots 
 \wh B^{(1)}_{a^1_{M_1}}(u_{1M_1})\,\wh F^{(1)}_{a^1_1\dots 
a^1_{M_1}}(\{u\})\,\Omega, \\
\wh F^{(1)}_{a^1_{1}\dots a^1_{M_{1}}}(\{u\}) &\in&
 (\CC^{\fm-1|\fn})^{\otimes M_1} \otimes \fD_{\fm-1|\fn},
\een
where $\wh F^{(1)}_{a^1_1\dots a^1_{M_{1}}}(\{u\})$ is built  
from operators $\wh d_{ij}(u)$, $2 \leq i \leq j \leq \fm+\fn$ only.
Since $\wh B^{(1)}(u)$ belongs to $\CC^{\fm-1|\fn}\otimes 
\fD_{\fm|\fn}$, 
we have introduced in the construction $M_{1}$ additional auxiliary 
spaces 
(labelled $a^1_{1},\ldots,a^1_{M_{1}}$) that are also carried by 
$\wh F^{(1)}_{a^1_1\dots a^1_{M_{1}}}(\{u\})$. These new auxiliary spaces 
take 
care of the linear combination one has to do between the different 
generators $\wh{d}_{1j}(u)$, $j=2,\ldots,\fm+\fn$, that enter into the 
construction.

Since $F^{(1)}_{a_1\dots a_{M_{1}}}(\{u\})$ is built up from 
operators $\wh d_{ij}(u)$, 
$2 \leq i \leq j \leq \fm+\fn$, it obeys the relation (proven in a 
more general context in theorem \ref{theo:embedding}):
\ben
\wh d_{11}(u)\, \wh F^{(1)}_{a_1\dots a_{M_1}}(\{u\})\,\Omega &=&
\wh \Lambda_1 (u)\, \wh F^{(1)}_{a_1\dots a_{M_1}}(\{u\})\,\Omega. 
\een
The transfer matrix is decomposed into
\ben
d(u^{(1)}) &=&\wt\fm_1(u)\,\wh d_{11}(u)+ \wh 
d^{(2)}(u)\mb{with}
\wh d^{(2)}(u) =  str_a\Big(M^{(2)}_a\,\wh D_a^{(2)}(u)\Big)\,\\
\wt\fm_1(u)&=&(-1)^{[1]}m_1\,k(u)+str(M^{(2)})\psi_1(u^{(1)}).
\een
 The action of $\wh d_{11}(u)$ on 
$\Phi(\{u^{(1)}\})$ takes the form
\ben
&&\wh d_{11} (u)\, \Phi(\{u\}) =\wh 
\Lambda_1(u)\,\prod_{i=1}^{M_1} 
\ff_1(u,u_{1i})
\,\Phi(\{u\})
+ \sum_{j=1}^{M_1} M_j(u;\{u_1\})\,\wh \Lambda_1(u_{1j})
\,\wh \Phi_j(u,\{u\})
\nonu
&&+ \sum_{j=1}^{M_1} N_j(u;\{u_1\})\,
\wh B^{(1)}_{a_1}(u_{11}) \dots \, \wh B^{(1)}_{a_j}(u)
 \dots \wh B^{(1)}_{a_{M_1}}(u_{1M_1}) 
 \wt d^{(2)}(u_{1j};\{u_1\})\,
\wh F^{(1)}_{a_1\dots a_{M_1}}(\{u\})\,\Omega,
\qquad\label{eq:td11Phi}
\een
with
$$M_j(u;\{u_1\}) = 
\fg_{1}(u,u_{1j}) 
\prod_{i \neq j}^{M_1} \ff_1(u_{1i},u_{1j})\mb{and} \,
N_j(u;\{u\}) = 
\frac{\fh_{1}(u,u_{1j})}{\fe_2(u_{1j})} 
\prod_{i \neq j}^{M_1} \wt \ff_2(u_{1i},u_{1j}).$$

The action of $\wh d^{(2)}(u)$ on $\Phi(\{u\})$ takes the form
\ben
&&\wh d^{(2)}(u)\,\Phi(\{u^{(1)}\}) \ =\  \prod_{j=1}^{M_1}
\wt \ff_2(u,u_{1j})\ B^{(1)}_{a_1}(u_{11}) \dots 
\wh B^{(1)}_{a_{M_1}}(u_{1M_1})\,\wt d^{(2)}(u;\{u_1\})\,
\wh F^{(1)}_{a_1\dots a_{M_1}}(\{u\})\,\Omega \nonu
&&\qquad+ \sum_{j=1}^{M_1} 
P_j(u;\{u_1\})\,\wh B^{(1)}_{a_1}(u_{11}) \dots 
B^{(1)}_{a_{j}} (u)\dots \wh
B^{(1)}_{a_{M_1}}(u_{1M_1})\,\wt d^{(2)}(u_{1j};\{u_1\})\,
\wh F^{(1)}_{a_1\dots a_{M_1}}(\{u\})\,\Omega \nonu
&&\qquad+ \sum_{j=1}^{M_1} 
Q_j(u;\{u_1\})\,\wh \Lambda_{1}(u_{1j})\,
\Phi_j(\{u\}), 
\een
with 
$$P_j(u;\{u_1\})=
\frac{\wt \fg_2(u,u_{1i})\,\fe_2(u)}{\fe_2(u_{1i})}
\prod_{i \neq 
j}^{M_1} \wt \ff_2(u_{1j},u_{1i}) \mb{;} \,
Q_j(u;\{u_1\})=\wt \fh_2(u,u_{1i})\,\fe_2(u)
\prod_{i \neq 
j}^{M_1} \ff_1(u_{1j},u_{1i})
.$$
where $\Phi_j(\{u\})$ is deduced from $\Phi(\{u\})$ 
by the change $u_{1j}\to\,u$. These expressions are
 computed as it has been done in section \ref{sec:ABA}: 
$N_1(u;\{u_1\})$, $M_1(u;\{u_1\})$,
$P_1(u;\{u_1\})$ and $Q_1(u;\{u_1\})$
are easy to compute; the other terms are obtained 
through a reordering of the operators $\wh B^{(1)}(u_{1j})$,
using the reordering lemma 
\ref{REOD} and the Yang-Baxter equation.
We also used the notation:
\ben
 \wt d^{(2)}(u;\{u_1\}) &=& 
 str_{a}\Big(M_a^{(2)}
 \prod_{j=1}^{\atopn{\longrightarrow}{M_1}}
 \wb{ \RR}_{a ,a_j}^{(2)}(u,u_{1j})
 \wh D_{a}^{(2)}(u) 
\prod_{j=1}^{\atopn{\longleftarrow}{M_1}}
 \RR_{a_j,a}^{(2)}(u,u_{1j})\Big) \,.
\een
\begin{rmk}\label{rmk:NABA2}
The fact that the wanted and unwanted terms contain the same operator $\wt 
d^{(2)}$ (but at different values $u$ and $u_{1j}$) allows 
to continue the nesting. 
In this way, the diagonalisation of this operator allows at the same 
time to compute the eigenvalue \textit{and} to show that the unwanted terms 
cancel (when the Bethe ansatz equations are obeyed). The apparition of 
this operator in the unwanted terms is directly related to the present 
form of the
commutation relations (\ref{eq:comdkkB}) and (\ref{eq:comDB}), see 
remark \ref{rmk:NABA}. Hence the need of a NABA couple to 
perform the nesting.
% 
% If one considers the wanted term only and asks for analyticity of the eigenvalues, 
% one gets the eigenvalues and BAEs for a 
% general diagonal matrix. This is the point of view adopted in 
% \cite{DVGR,GM,YZ}. However, one cannot show in this way that the unwanted 
% terms cancel, and one does not get the form of the Bethe vectors. 
\end{rmk}

As already mentionned, the calculation makes appear a new transfer 
matrix $\wt 
d^{(2)}(u;\{u_1\})$ corresponding to a $\fD_{\fm-1|\fn}$ chain with 
$L+M_{1}$ sites, 
the $M_{1}$
additional sites corresponding to fundamental representations of 
$\fD_{\fm-1|\fn}$. This interpretation is supported by theorem
\ref{theo:embedding}
which ensure that $\wt{D}_{a}^{(2)}(u;\{u_1\})$ generates 
$\fD_{\fm-1|\fn}$, and that $\wt d^{(2)}(u;\{u_1\})$ is indeed the 
transfer matrix of an integrable spin chain.
Then, if we assume that $F^{(1)}_{a_1\dots a_{M_1}}(\{u\})\,\Omega$ 
is an eigenvector of this new transfer matrix,
\ben
\wt d^{(2)}(u;\{u_1\})\,\wh F^{(1)}_{a_1\dots a_{M_1}}(\{u\})\,\Omega 
&=& 
\wt\Gamma^{(2)}(u)\,\wh F^{(1)}_{a_1\dots a_{M_1}}(\{u\})\,\Omega\,,
\label{NPB}
\een
we deduce:
\ben
\wh d_{11} (u)\, \Phi(\{u\}) &=&\wh 
\Lambda_1(u)\,\prod_{i=1}^{M_1} 
\ff_1(u,u_{1i})
\,\Phi(\{u\})
\nonu
&& + \sum_{j=1}^{M_1}\Big( M_j(u;\{u_1\})\,
\wh  \Lambda_1(u_{1j})+N_j(u;\{u_1\})
\,\wt \Gamma^{(2)}(u_{1j})\Big)\, \Phi_j(\{u\}), 
\label{eq:d11Phif} 
\\
\wh d^{(2)}(u)\,\Phi(\{u\}) &=& \wt\Gamma^{(2)}(u)\, 
\prod_{j=1}^{M_1}\wt \ff_2(u,u_{1j})\,\Phi(\{u\})
\nonu
&&+ 
\sum_{j=1}^{M_1}\Big( 
P_j(u;\{u_1\})\,\wt \Gamma^{(2)}(u_{1j})+
Q_j(u;\{u_1\})\,\wh \Lambda_1(u_{1j})\Big)\,
\Phi_j(\{u\}).
\label{eq:d2Phif} 
\een
Gathering these relations together, we get a first 
expression of the action of $d(u)$ on $\Phi (\{u\})$.
When we cancel in this expression the unwanted terms (carried by 
$\Phi_{j}(\{u\})$), we get the first system of Bethe 
equations:
\ben
\frac{\wh \Lambda_1(u_{1j})}{\wt\Gamma^{(2)} (u_{1j})}=
\frac{\chi_1(u_{1j})}{\fe_2(u_{1j})}\prod_{i\neq j}^{M_1}
\frac{\wt 
\ff_2(u_{1j},u_{1i})}{\ff_1(u_{1j},u_{1i})}.
\een
We also get a first expression of the transfer matrix eigenvalue:
\ben
\wh{d}^{(1)}(u)\,\Phi (\{u\}) &=& 
\left(\,\wt\fm_1(u)\, 
\wh \Lambda_1(u)\prod_{j=1}^{M_1}\ff_1(u,u_{1j})+
\wt\Gamma^{(2)}(u)\prod_{j=1}^{M_1} 
\wt \ff_1(u,u_{1j})
\right)\, \Phi(\{u\})\,.\qquad
\een
In the above relations, everything is known \textit{but} the 
eigenvalue $\wt \Gamma^{(2)}(u)$, introduced in (\ref{NPB}), and the 
explicit form of $\wh{F}^{(1)}_{a^1_1\dots a^1_{M_1}}(\{u\})$ ensuring 
that 
(\ref{NPB}) is indeed satisfied.

Thus, at the end of this first recursion step, 
we have `reduced' the problem of 
computing an eigenvector $\Phi(\{u\})$ for the transfer matrix $d(u)$ 
of a $\fD_{\fm|\fn}$ chain with $L$ sites to the problem of 
computing 
an eigenvector $\Phi^{(1)}(\{u\})=
\wh{F}^{(1)}_{a^1_1\dots a^1_{M_1}}(\{u\})\,\Omega$ for the
transfer matrix $\wt d^{(2)}(u;\{u_1\})$ 
of a $\fD_{\fm-1|\fn}$ chain with $L+M_{1}$ sites.

\null
\begin{rmk}[Change of notation] \label{rmk:change}
To avoid too complicated notation for 
the second step, we need to slightly change the notation at the end of 
the first step. First, we rename the hatted operators $\wh X(u)\ \to\ 
X(u)$, although they still have the spectral parameter shift and the 
operator transformation coming from the first step. Secondly, we omit 
the tilda on operators, $\wt X (u) \to X(u)$, keeping in mind that the 
new operator $X(u)$ have $M_1$ sites more than the one of the previous step.
In this way, we will be able to re-use the `hatted' and `tilded' 
notations for the transformations used in the second step. 

This will 
be the general approach at each step: at the end of step $k$, we will 
perform a change of notation, suppressing the hats and tildas on 
operators, to use them again in step $k+1$.
\end{rmk}

It remains to single out the highest 
weights corresponding to the fundamental representations carried by 
the new sites. This is done in the following way  
\ben
\Phi^{(1)}(\{u\}) &=& F^{(1)}_{a_1\dots 
a_{M_1}}(\{u\})\,\Omega, \nonu
\Phi^{(1)}(\{u\}) &=& 
B^{(2)}_{a^2_1} (u_{21};\{u_{1}\}) \dots 
B^{(2)}_{a^2_{M_{2}}} (u_{2\,M_{2}};\{u_1\})\,
F^{(2)}_{a^2_1 \dots a^2_{M^2}}(\{u\})\,\Omega^{(2)}, \\
\Omega^{(2)} &=& (e^{(1)}_1)^{\otimes^{M_1}} \otimes \Omega,\,
\een
where $e^{(1)}_1=(1,0,\ldots,0)^t \in \CC^{\fm-1|\fn}$ and 
$F^{(2)}_{a^{2}_1\dots 
a^{2}_{M_{2}}}(\{u\})$ is built on operators $
d_{ij}(u^{(2)};\{u_1\})$, with $j\geq i>2$. The 
operators $ B^{(2)}(u;\{u_1\})$ play the role, for the 
$\fD_{\fm-1|\fn}$ chain of length $L+M_{1}$, of the operators 
$B^{(1)}(u)$ for the $\fD_{\fm|\fn}$ chain of length $L$. Explicitly, 
they are obtained from the decomposition (\ref{eq:mono-k}) of the monodromy 
matrix.
 
%%%%%%%%%%%%%%%%%%%%%%%%%%%%%%%%%%%%
\subsection{General step}

More generally, the step $k$ starts with the problem
\begin{equation}
 d^{(k)}(u;\{u_{k-1}\})\,\Phi^{(k-1)}(\{u\})
= \Gamma^{(k)}(u)\,\Phi^{(k-1)}(\{u\}),
\label{eq:tkPhik}
\end{equation}
where 
$ d^{(k)}(u;\{u_{k-1}\}) = str\Big(M^{(k)} 
D^{(k)}(u;\{u_{k-1}\})\Big)$
 is the transfer matrix of a $\fD_{\fm-k-1|\fn}$ spin chain of length 
$L+\sum_{j=1}^{k-1} M_{j}$ (obtained from the previous step).
We recall that hats and tildas have been suppressed, according to 
remark \ref{rmk:change}, including for the function $\Gamma^{(k)}(u)$.
We define
\ben
\Phi^{(k-1)}(\{u\}) &=& F^{(k-1)}_{ a^{k-1}_1 \dots a^{k-1}_{M_{k-1}} 
}
(\{u\})\,\Omega^{(k-1)}
\ =\ \BB^{(k)}(\{u_k\})
\,F^{(k)}_{a^{k}_1\dots a^{k}_{M_{k}}}(\{u\})\,\Omega^{(k)},
\\
\Omega^{(k)} &=& (e^{(k-1)}_1)^{\otimes{M_{k-1}}} \,\otimes 
\,\Omega^{(k-1)}\,,
\een
with $e^{(k)}_1=(1,0,\ldots,0)^t \in \CC^{\fm-k|\fn}$.
We have introduced
\ben
\BB^{(k)}(\{u_{k}\}) &=& B^{(k)}_{a^{k}_1}(u_{k1};\{u_{k-1}\}) 
\dots 
 B^{(k)}_{a^{k}_{M_{k}}}(u_{kM_{k}};\{u_{k-1}\}),
 \label{eq:BB}
\een
where the operators are extracted from the monodromy matrix, see 
equation (\ref{eq:mono-k}).

\begin{rmk}
\label{rmk:auxspace}
In (\ref{eq:BB}), we have indicated only the auxiliary spaces 
$a^k_{j}$, $j=1,\ldots,M_{k}$. In fact, since $ 
D^{(k)}(u;\{u_{k-1}\})$ is viewed 
as the monodromy matrix of a spin chain of length 
$L+\sum_{j=1}^{k-1}M_{j}$, the other spaces $a^\ell_{j}$, 
$j=1,\ldots,M_{\ell}$, $\ell<k$, are now quantum spaces. Thus, they 
do not appear explicitly in $ D^{(k)}$, as the sites of the 
original spin chain, but obviously this monodromy matrix (and its 
components) does depend on all these spaces.
\end{rmk}
We extract from $ d^{(k)}(u;\{u_{k-1}\})$ the component 
$d_{kk}(u;\{u_{k-1}\})$:
\beq
d^{(k)}(u;\{u_{k-1}\}) = (-1)^{[k]}\,m_k\,d_{kk}
(u;\{u_{k-1}\})
+str\Big(M^{(k+1)}D^{(k+1)} (u;\{u_{k-1}\}) \Big).\,
\eeq
Now we must transform the operator: 
\ben
d_{kk}(u^{(k)})&=&\wh d_{kk}(u)\,,\qquad
B^{(k)}_a(u^{(k)})=\wh B^{(k)}_a(u), 
\nonu
D^{(k+1)}(u^{(k)})&=&\wh D^{(k+1)}(u)+
\psi_k(u^{(k)})\, \wh d_{kk}(u)\, \II^{(k)}. 
\een
The transfer matrix $d^{(k)}(u^{(k)};\{u_{k-1}\})$ is rewritten as
\ben
\wh d^{(k)}(u;\{u_{k-1}\}) &=&\,\wt\fm_k(u)
\,\wh{d}_{kk}(u;\{u_{k-1}\})
+str\Big(M^{(k+1)}\wh{D}^{(k+1)} (u;\{u_{k-1}\}) \Big)
\,,\\
\wt\fm_k(u)&=&(-1)^{[k]}m_k\,+str(M^{(k)})\psi_k(u^{(k)}),
\een
and the Bethe vector: 
\ben
\Phi^{(k-1)}(\{u\}) &=& 
\wh \BB^{(k)}(\{u_{k}\})\,\wh F^{(k)}_{a^{k}_1\dots 
a^{k}_{M_{k}}}(\{u\})\,\Omega^{(k)},
\\
\wh \BB^{(k)}(\{u_{k}\}) &=& \wh 
B^{(k)}_{a^{k}_1}(u_{k1};\{u_{k-1}\}) \dots 
\wh B^{(k)}_{a^{k}_{M_{k}}}(u_{kM_{k}};\{u_{k-1}\}).
\een
Now we can compute the action of the transfer matrix on this vector. 
We first commute
 $\wh d_{kk}(u;\{u_{k-1}\})$ 
 and 
$\wh d^{(k+1)} (u;\{u_{k-1}\})=
str\Big(M^{(k)}\wh{D}^{(k+1)} (u;\{u_{k-1}\})\Big)$
 with the operator 
$\wh \BB^{(k)}(\{u_{k}\})$:
\ben
&&\wh d_{kk}(u;\{u_{k-1}\})\, \Phi^{(k-1)}(\{u^{(k)}\}) \ =\ 
\prod_{j=1}^{M_k} 
\ff_k(u,u_{kj})\ \wh \BB^{(k)}(\{u_{k}\})\ 
\wh d_{kk}(u;\{u_{k-1}\})\, \Phi^{(k)}(\{u\}) 
\nonu
&&\qquad\qquad + \sum_{j=1}^{M_k} M_j(u;\{u_{k-1}\})\ 
\wh \BB^{(k)}_j(u;\{u_{k}\}) 
\wh d_{kk}(u_{kj};\{u_{k-1}\})\, \Phi^{(k)}(\{u\}) 
\nonu
&& \qquad \qquad + \sum_{j=1}^{M_k} N_j(u;\{u_{k-1}\})\ 
\wh \BB^{(k)}_j(u;\{u_{k}\}) 
\wt d^{(k+1)}(u_{kj};\{u_{k}\}) \, \Phi^{(k)}(\{u\}), 
\label{eq:dkkPhi}
\een
\ben
&& \wh d^{(k+1)}(u;\{u_{k-1}\}) \, \Phi^{(k-1)}(\{u\}) \ =\ 
\prod_{j=1}^{M_k} 
\wt \ff_{k+1}(u,u_{kj})\ \wh \BB^{(k)} (\{u_k\})\ 
\wt d^{(k+1)}(u;\{u_k\})\, \Phi^{(k)}(\{u\})
\nonu
&& \qquad\qquad + \sum_{j=1}^{M_k} P_j(u;\{u_{k-1}\})\ 
\wh \BB^{(k)}_j(u;\{u_k\}) 
\wt d^{(k+1)}(u_{kj};\{u_k\}) \, \Phi^{(k)}(\{u\})
\nonu
&& \qquad \qquad + \sum_{j=1}^{M_k} Q_j(u;\{u_{k-1}\})\ 
\wh\BB^{(k)}_j(u;\{u_{k}\}) 
\wh d_{kk}(u_{kj};\{u_{k-1}\}) \, \Phi^{(k)}(\{u\}), 
\label{eq:dkPhi}
\een
where we have introduced:
\ben
 \Phi^{(k)}(\{u\}) &=&\wh F^{(k)}_{a^{k}_1\dots 
a^{k}_{M_{k}}}(\{u\})\,\Omega^{(k)},
\nonu
\wt d^{(k+1)}(u;\{u_{k}\})&=&
str_a\left(M_a^{(k+1)}\,\Big(\prod_{j=1}^{\atopn{\longrightarrow}{M_k}} 
\wb{\RR}_{aa^k_j}^{(k+1)}(u,u_{kj})\Big)\,
\wh D_{a}^{(k+1)}(u;\{u_{k-1}\}) 
\,\Big(\prod_{j=1}^{\atopn{\longleftarrow}{M_k}} 
\RR_{a^k_ja}^{(k+1)}(u,u_{kj})\Big)\right).\nonumber
\een
The functions $M_{j},N_{j},P_{j}$ and $Q_{j}$ are the same as in the first step 
(section \ref{first}) but with indices 
$1 \to k$ on  functions and Bethe roots. 
We use 
the following reordering lemma:

\begin{lemma}
For each $k=1,\ldots,\fm+\fn-1$ and $j=1,\ldots,M_{k}$, we have  
\begin{eqnarray}
\wh \BB^{(k)}(\{u_k\}) &=& \wh B^{(k)}_{j}(u_{kj})\,\wh 
B^{(k)}_1(u_{k1}) \dots 
\wh B^{(k)}_{j-1}(u_{k,j-1})\,\wh B^{(k)}_{j+1}(u_{k,j+1}) \dots
\wh B^{(k)}_{M_{k}}(u_{kM_{k}})\nonu
&&\times\prod_{i=1}^{\atopn{\longrightarrow}{j-1}}(-1)^{[j]}\,
\frac{\fa_{j+1}(u^{(k)}_{ki},u^{(k)}_{kj})}
{\fa_{j}(u^{(k)}_{ki},u^{(k)}_{kj})}\,
\RR_{ji}^{(k+1)}(u_{ki},u_{kj})
\end{eqnarray} 
where the dependence in $\{u_{k-1}\}$ has been omitted in $\wh 
B^{(k)}_{p}$.
\label{REOD}
\end{lemma}
\prf
 Direct calculation using the commutation relations 
(\ref{eq:comBB})-(\ref{eq:comDB}).   
\finprf

We now compute the action of $\wh d_{kk}(u ;\{u_{k-1}\})$ 
and $\wt d^{(k+1)}(u;\{u_k\})$ on 
$\wh F^{(k)}(\{u\})\,\Omega^{(k)}$.
These actions follow from the theorem \ref{theo:embedding}.
For $\wh d_{kk}(u ;\{u_{k-1}\})$  we have:
\ben
\wh d_{kk}(u ;\{u_{k-1}\})\wh F^{(k)}(\{u\})\,\Omega^{(k)}=\wh 
\Lambda_{k}(u ;\{u_{k-1}\})\wh F^{(k)}(\{u\})\,\Omega^{(k)}.
\een
It remains to do the same for 
$\wt d^{(k+1)}(u;\{u_k\})$. 
It corresponds to a new monodromy matrix
\ben
\wt D_{a}^{(k+1)}(u;\{u_k\}) &=& 
 \prod_{j=1}^{\atopn{\longrightarrow}{M_k}} 
\wb{\RR}_{aa^k_j}^{(k+1)}(u,u_{kj})
\wh D_{a}^{(k+1)}(u;\{u_{k-1}\}) 
\prod_{j=1}^{\atopn{\longleftarrow}{M_k}} 
\RR_{a^k_ja}^{(k+1)}(u,u_{kj}) \,.
\een
It also satisfies the reflexion equation,
see the theorem \ref{theo:embedding},
so that the problem 
is integrable, and defines a $\fD_{\fm-k|\fn}$ spin chain, with 
$L+\sum_{j=1}^{k}M_{j}$ sites.

We get a new eigenvalue problem:
\begin{equation}
\wt d^{(k+1)}(u;\{u_k\}) \,\Phi^{(k)}(\{u\}) 
= \wt\Gamma^{(k+1)}(u)\,\Phi^{(k)}(\{u\}). 
\label{eq:chain-k}
\end{equation}
Assuming the form (\ref{eq:chain-k}), we can show, following the same 
lines as in the first step, that $\Phi^{(k-1)}(\{u\})$ is a transfer 
matrix 
eigenvector provided the  
the $k^{th}$ system of Bethe equations,
\ben
\frac{\wh\Lambda_k(u_{kj};\{u_{k-1}\})}{\wt\Gamma^{(k+1)}(u_{kj};\{u_{k}\})}&=&
\frac{\chi_k(u_{kj})}{\fe_{k+1}(u_{kj})}
\prod_{i\neq j}^{M_k} 
\frac{\wt{\ff}_{k+1}(u_{kj},u_{ki})}
{\ff_k(u_{kj},u_{ki})},
\label{eq:protoBE}
\een
is obeyed. We also get an expression for $\wh\Gamma^{(k)}(u)$, the eigenvalue of 
$\wh d^{(k)}(u)$:
\beq
\wh \Gamma^{(k)}(u) = \,\wt\fm_k(u)
\, \prod_{j=1}^{M_k} 
\ff_k(u,u_{kj})\wh \Lambda_{k}(u;\{u_{k-1}\})\,
+\prod_{j=1}^{M_k} 
\wt \ff_{k+1}(u,u_{kj})\ \wt \Gamma^{(k+1)}(u;\{u_{k}\}).
\eeq

%%%%%%%%%%%%%%%%%%%%%%%%%%%%%%%%%%%%
\subsection{End of induction}

To end the recursion, we use the $\fm+\fn=2$ case and remark that:
\ben
\wh \Gamma^{(\fn+\fm)}(u)=\wh \Lambda_{\fn+\fm}(u,\{u_{\fm+\fn-2}\}).
\label{eq:gamma-fin}
\een
Using the shift notation,
$u^{(k\dots l)}=(\dots (u^{(k)})^{(k+1)}\dots)^{(l)}$, for $k\leq l$,
we deduce from (\ref{eq:gamma-fin})
that $\wh\Gamma$ is expressed in term of $\wh\Lambda$:
\begin{eqnarray}
&&\wh\Gamma^{(k+1)}(u^{(k+2 \dots \fn+\fm-1)}) \ =\ 
\,\wt\fm_{k+1}(u^{(k+2 \dots \fn+\fm-1)})
\,\wh \Lambda_{k+1}(u^{^{(k+2 \dots \fn+\fm-1)}};\{u_{k}\})\,
\cF_{k+1}(u)\qquad
\nonumber
\\
&&\qquad+ \sum_{\ell=k+2}^{\fm+\fn-1}
\,\wt\fm_\ell(u^{(\ell+1 \dots \fn+\fm-1)})
\,\wh \Lambda_{\ell}(u^{(\ell+1 \dots \fn+\fm-1)};\{u_{\ell-1}\})
\ \cF_{\ell}(u)\ \prod_{p=k+1}^{\ell-1}\,\wt \cF_{p}(u)
\nonu
&&\qquad+(-1)^{[\fm+\fn]} m_{\fm+\fn}
\,\wh\Lambda_{\fm+\fn}(u;\{u_{\fm+\fn-2}\})
\prod_{p=k+1}^{\fm+\fn-1}\,\wt \cF_{p}(u),
\een
where we have introduced
\beq
 \cF_{\ell}(u) = \prod_{j=1}^{M_{\ell}}\ff_{\ell}(u^{(\ell+1 
\dots \fn+\fm-1)},u_{\ell j})\,
,\quad
\wt\cF_{\ell}(u) = \prod_{j=1}^{M_{\ell}}\wt\ff_{\ell+1}(u^{(\ell+1 
\dots \fn+\fm-1)},u_{\ell j})\,, \quad
 \ell \in \{k \dots \fm+\fn-1\} \,,
\nonumber
\eeq
with the convention $u^{(k\dots l)}=u$ if $k>l$.
It remains to compute the values 
$\wh \Lambda_k(u;\{u_{k-1}\})$:

\begin{lemma}\label{lem:lambdatilde}
The eigenvalue $\wh\Lambda_k(u;\{u_{k-1}\})$ of 
$\wh d_{kk}(u;\{u_{k-1}\})$ on $\Omega^{(k-1)}$ is given by,
\begin{equation}
\wh\Lambda_k(u;\{u_k\}) =\wh \Lambda_k(u)\,
\prod_{\ell=1}^{k-2}\ \prod_{j=1}^{M_{\ell}} 
\frac{1}
{\wt \ff_{\ell+1}(u^{(\ell+1 \dots k)},u_{\ell j})}
=\wh \Lambda_k(u)\,
\prod_{\ell=1}^{k-2}\  
\frac{1}{\wt\cF_{\ell}(u^{\wb{(k+1 \dots \fm+\fn-1)}})}\,, 
\quad k=1,\ldots,\fm+\fn,
\label{eq:lambtil}
\end{equation}
where we have used $\wh d_{kk}(u)\,\Omega=\wh \Lambda_k(u)\,\Omega$
with:
\ben
\wh \Lambda_k(u)&=&\Lambda_k(u^{(1\dots k)})-\sum_{i=1}^{k-1} 
\,q^{2(k-1-i)-4\sum_{l=i+1}^{k-1}[l]}\,\psi_i((u^{(k)})^{(i)}) 
\,\Lambda_i(u^{(1\dots k)})
\label{eq:lambhat1}
 \\
 &=&
\cK_{k}(u^{(1\dots k)})\,\lambda_{k}(u^{(1\dots k)})\,
\lambda'_{k}(u^{(1\dots k)}).
\label{eq:lambhat}
\een
\end{lemma}
\prf
First we introduce a useful property between coproduct and supertrace:
\ben
str_a(\Delta(D^{(k)}_a(u)) =\Delta(str_a(D^{(k)}_a(u))).
\een
It is obvious because supertrace and  coproduct do not act in the same 
space.
We recall the fundamental representation evaluation map
for the $\cA_{\fm-k+1|\fn}$ algebra:
\ben
\pi^{(k)}_a: \begin{array}{lclcl}   
\cA_{\fm-k+1|\fn}\otimes End(\CC^{\fm|\fn}) & \to &  
End(\CC^{\fm-k+1|\fn})\otimes End(\CC^{\fm|\fn})\\[.21ex]
T^{(k)}(u) & \mapsto & \displaystyle \,\wb 
\RR^{(k)}_{12}(u,a)\\[.21ex]
(T^{-1}(\inv(u)))^{(k)} & \mapsto & \displaystyle \, 
\RR^{(k)}_{21}(u,a) 
\end{array}\,.
\label{eq:eval-k}
\een
We also need the representation of $\cA_{\fm-k+1|\fn}$ induced by 
the inclusion $\cA_{\fm-k+1|\fn}\hookrightarrow \cA_{\fm-p+1|\fn}$, 
$p<k$. From the identity
\ben
T^{(k)}(u)=\II^{(k)}\, (T^{(p)}(u^{(p+1 \dots k)}))\, \II^{(k)}\,,
\een
 we can deduce the form of  $\pi^{(p)}_v ((T^{(k)}(u))$ in the 
 fundamental representation of $\cA_{\fm-p+1|\fn}$:
 \ben
(id  \otimes \pi^{(p)}_v) (T^{(k)}(u))&=&\II^{(k)}_1 (id  \otimes 
\pi^{(p)}_v) (T^{(p)}(u^{(p+1 \dots k)})) \II^{(k)}_1 
=\II^{(k)}_1\,\wb{\RR}^{(p)}_{12}(u^{(p+1 \dots k)},v)\,\II^{(k)}_1 
\nonu
&=&\wb{\RR}^{(k,p)}_{a,b}(u^{(p+1 \dots k)},v).
\een
The last equality is just the definition of 
$\RR^{(k,p)}_{a,b}(u^{(p+1 \dots k)},v)$, see (\ref{eq:Rpk}).

Hence, using theorem \ref{embedding-co}, we can rewrite the monodromy operator $\wt 
D_{a}^{(k+1)}(u;\{u_{k-1}\})$ as:
\ben
\wt 
D_{a}^{(k+1)}(u;\{u_{k-1}\})&=&
\Big(\prod_{j=1}^{\atopn{\longrightarrow}{M_k}} 
\wb{\RR}_{aa^k_j}^{(k+1)}(u,u_{kj})\Big)\,
\wh D_{a}^{(k+1)}(u;\{u_{k-1}\}) 
\,\Big(\prod_{j=1}^{\atopn{\longleftarrow}{M_k}} 
\RR_{a^k_ja}^{(k+1)}(u,u_{kj})\Big) \nonu
&=&(id  \otimes (\pi^{(k+1)}_{u_{i k}})^{\otimes^{M_k}_{i=1}}) \circ 
\Delta^{M_k} (\wh D_{a}^{(k+1)}(u;\{u_{k-1}\}))\,,
\een
while the operator $  \wt {d}^{(k)}_{kk}(u) $
takes the form:
\ben
\wt d^{(k)}_{kk}(u;\{u_{k-1}\})=(id  \otimes ( 
(\pi^{(p+1)}_{u_{pi}})^{\otimes^{M_p}_{i=1}}) )^{\otimes_{p=1}^{k-1}}) 
\circ \Delta^{\sum_{p=1}^{k}\,M_p}\,\big(\wh{d}^{(k)}_{kk}(u)\big).
\een
Now acting on the highest weight $\Omega^{(k-1)}$ and using lemma 
\ref{lem-1}, we find the following result:
\ben
\wt d^{(k)}_{kk}(u)\,\Omega^{(k-1)}= (id  \otimes ( (\pi^{(p+1)}_{u_{i 
p}})^{\otimes^{M_p}_{i=1}}) )^{\otimes_{p=1}^{k}}) (t^{(k)}_{kk}(u) 
t'^{(k)}_{kk}(\inv(u)))^{\otimes^{\sum_{p=1}^{k}\,M_p}} \otimes 
\wh{d}^{(k)}_{kk}(u)\,\Omega^{(k-1)}.
\een
We have from the definition of $\RR$ matrices and $\Omega^{(k-1)}$,
\ben
\pi^{(p+1)}_{u_{i p}} \big(t^{(k)}_{kk}(u)\, t'^{(k)}_{kk}(\inv(u))\big)\,
\Omega^{(k-1)} &=&
\frac{\fb(u^{(p+1\dots k)},u_{ip})\, \bar{\fb}(u^{(p+1\dots 
k)},u_{ip})}{\fa_{p+1}(u^{(p+1\dots k)},u_{ip}) \,
\bar{\fa}_{p+1}(u^{(p+1\dots k)},u_{i p})}\, \Omega^{(k-1)}, \nonu
\pi^{(k)}_{u_{i k}} \big(t^{(k)}_{kk}(u)\, t'^{(k)}_{kk}(\inv(u))
\big)\, \Omega^{(k-1)} 
&=&\Omega^{(k-1)},
\een
that leads to the result (\ref{eq:lambtil}). The eigenvalue 
(\ref{eq:lambhat1}) is computed directly from theorem 
\ref{theo:embedding}. To obtain the form (\ref{eq:lambhat}), one uses 
the equalities (\ref{eq:valprD}), (\ref{eq:grKapp}) and the identity 
(\ref{eq:id-psi}).
\finprf

{From} the expression given in lemma \ref{lem:lambdatilde}, one 
deduces that:
\ben
\label{eq:gammak}
\wh \Gamma^{(k+1)}(u^{(k+2\dots \fn+\fm-1)}) &=& \left\{
\,\wt\fm_{k+1}(u^{(k+2\dots \fn+\fm-1)})\,\wh 
\Lambda_{k+1}(u^{(k+2\dots \fn+\fm-1)})\,
\cF_{k+1}(u)\right.
\nonu
&& + \sum_{\ell=k+2}^{\fm+\fn-1}
\,\wt\fm_\ell(u^{(\ell+1 \dots \fn+\fm-1)})
\, \wh \Lambda_{\ell}(u^{(\ell+1 \dots \fn+\fm-1)})\,
\cF_l(u) \, \wt\cF_{l-1}(u)\nonu
&&\left. +(-1)^{[\fm+\fn]} m_{\fm+\fn}
\,\wh \Lambda_{\fm+\fn}(u^{(\fn+\fm-1)})\,
\wt\cF_{\fm+\fn-1}(u)\rule{0ex}{2.1ex}\right\}\,\frac{1}{\prod_{\ell=1}^{k-1}\wt\cF_{\ell}(u)}\,.
\een
Let us note that since $\fb(u,u)=0$, equation 
(\ref{eq:gammak}) implies that:
\begin{eqnarray}
\wh \Gamma^{(k+1)}(u_{\ell j}) &=& 0,
\mb{for} j=1,\ldots,M_{\ell}\,;\ \ell=1,\ldots,k-1,
\label{eq:gam-part1}\\
\wh \Gamma^{(k+1)}(u_{kj}) &=& 
\wt\fm_{k+1}(u_{kj})\,\wh \Lambda_{k+1}(u_{kj})
\cF_{k+1}(u_{kj}^{\wb{(k+2\ldots\fn+\fm-1)}})
\prod_{\ell=1}^{k-1}\ 
\frac{1}
{\wt{\cF}_{\ell+1}(u_{kj}^{\wb{(k+2\ldots\fn+\fm-1)}})} 
, \nonu
&& \mb{for}j=1,\ldots,M_{k}\,.
\label{eq:gam-part2}
\end{eqnarray}

%%%%%%%%%%%%%%%%%%%%%%%%%%%%%%%%%%%%
\subsection{Final form of Bethe vectors, eigenvalues and equations}
Using expressions (\ref{eq:gam-part1}), (\ref{eq:gam-part2}), and the value of 
$\wh \Lambda_{k}(u;\{u^{(k)}\})$ given in lemma \ref{lem:lambdatilde}, 
one can recast the Bethe equations (\ref{eq:protoBE}) in their final form:
\ben
\frac{\wh \Lambda_{k} (u_{kj}) }{\wh \Lambda_{k+1} (u_{kj}^{\wb{(k+1)}}) } &=& 
\frac{\wt\fm_{k+1}(u_{kj}^{\wb{(k+1)}})\,\chi_k(u_{kj})}{\,\fe_{k+1}(u_{kj})}
\prod_{i=1}^{M_{k-1}}\frac{1}{\wt\ff_{k}(u_{k j}^{(k)},u_{k-1,i})}
\prod_{i\neq j}^{M_k} 
\frac{\wt{\ff}_{k+1}(u_{kj},u_{ki})}{\ff_k(u_{kj},u_{ki})}
\prod_{i=1}^{M_{k+1}} 
{\ff}_{k+1}(u_{k j}^{\wb{(k+1)}},u_{k+1,i}),
\nonu
&& j=1,\ldots,M_{k}\,,\quad k=1,\ldots,\fm+\fn-1,
\label{BE}
\een
with the convention $M_{0}=M_{\fm+\fn}=0$.
The number of  parameter families is $\fm+\fn-1$.
We checked that using the weights (\ref{eq:Lambda-eval}), (\ref{VP}) and the functions 
given in appendix \ref{app:fct}, one reproduces the BAEs already 
computed, in e.g. \cite{KR,reshe,DVGR}, and also the general forms given 
in \cite{ACDFR,ACDFR2,RS}. In particular, for the  
fundamental weight 
$\mu=(1,0,\dots,0)$, we recovers the BAEs for a spin chain with 
fundamental representations. For instance, for the case of 
$\wh\cU_{q}(2|2)$, 
which may be of some relevance in the context of AdS/CFT 
correspondence, one gets (for $L$ sites with evaluation parameter $b_i$
 , $a_+=1$ and $a_-=2$):

{\footnotesize
\begin{eqnarray} 
&&\frac{c_+\,u_{1i}\,q^{\frac{1}{2}}-\frac{q^{-\frac{1}{2}}}{c_+\,u_{1i}}}
{\frac{u_{1i}}{c_+}\,q^{\frac{1}{2}}-\frac{c_+}{u_{1i}}q^{-\frac{1}{2}}}
\prod_{i=1}^{L}\frac{(\frac{u_{1i}}{b_i}q^{\frac{1}{2}}-\frac{b_i}{u_{1i}}q^{-\frac{1}{2}})
(u_{1i}\,b_i\,q^{\frac{1}{2}}-\frac{q^{-\frac{1}{2}}}{b_i\,u_{1i}})}
{(\frac{u_{1i}}{b_i}q^{-\frac{1}{2}}-\frac{b_i}{u_{1i}}q^{\frac{1}{2}})
(u_{1i}\,b_i\,q^{-\frac{1}{2}}-\frac{q^{\frac{1}{2}}}{b_i\,u_{1i}})}\nonu
&&\qquad= \prod_{j\neq i}^{M_{1}}
\frac{(\frac{u_{1i}}{u_{1j}}q-\frac{u_{1j}}{u_{1i}}q^{-1})
(u_{1i}\,u_{1j}\,q-\frac{q^{-1}}{u_{1i}\,u_{1j}})}
{(\frac{u_{1j}}{u_{1i}}q-\frac{u_{1i}}{u_{1j}}q^{-1})
(u_{1i}\,u_{1j}\,q^{-1}-\frac{q}{u_{1i}\,u_{1j}})}\ 
\prod_{j=1}^{M_{2}}
\frac{(\frac{u_{2j}}{u_{1i}}q^{-\frac{1}{2}}-\frac{u_{1i}}{u_{2j}}q^{\frac{1}{2}})
(u_{2j}\,u_{1i}\,q^{-\frac{1}{2}}-\frac{q^{\frac{1}{2}}}{u_{2j}\,u_{1i}})}
{(\frac{u_{2j}}{u_{1i}}q^{-\frac{1}{2}}-\frac{u_{2j}}{u_{1i}}q^{\frac{1}{2}})
(u_{2j}\,u_{1i}\,q^{\frac{1}{2}}-\frac{q^{-\frac{1}{2}}}{u_{2j}\,u_{1i}})}\ ,
\nonu
&& i=1,\ldots,M_{1}\nonumber
\een
\ben
\frac{u_{2i}^2\,q^{-1}-c^2_{-}\,q}{u_{2i}^{-2}\,-c^2_{-}q^2}(-1)^{M_2+1} &=& \prod_{j=1}^{M_{1}}\frac{(
\frac{u_{2i}}{u_{1j}}q^{-\frac{1}{2}}-\frac{u_{1j}}{u_{2i}}q^{\frac{1}{2}})
(u_{1j}\,u_{2i}\,q^{-\frac{1}{2}}-\frac{q^{\frac{1}{2}}}{u_{1j}\,u_{2i}})}
{(\frac{u_{2i}}{u_{1j}}q^{\frac{1}{2}}-\frac{u_{1j}}{u_{2i}}q^{-\frac{1}{2}})
(u_{1j}\,u_{2i}\,q^{\frac{1}{2}}-\frac{q^{-\frac{1}{2}}}{u_{1j}\,u_{2i}})}\ 
\prod_{j=1}^{M_{3}}
\frac{(\frac{u_{3j}}{u_{2i}}q^{-\frac{1}{2}}-\frac{u_{2i}}{u_{3j}}q^{\frac{1}{2}})
(u_{3j}\,u_{2i}\,q^\frac{1}{2}-\frac{q^{-\frac{1}{2}}}{u_{3j}\,u_{2i}})}{(
\frac{u_{3j}}{u_{2i}}q^{\frac{1}{2}}-\frac{u_{3j}}{u_{2i}}q^{-\frac{1}{2}})
(u_{3j}\,u_{2i}\,q^{-\frac{1}{2}}-\frac{q^\frac{1}{2}}{u_{3j}\,u_{2i}})}\ 
\nonu
i=1,\ldots,M_{2}\nonu
-1 &=& \prod_{j=1}^{M_{2}}
\frac{(\frac{u_{3i}}{u_{2j}}q^{\frac{1}{2}}-\frac{u_{2j}}{u_{3i}}q^{-\frac{1}{2}})
(u_{2j}\,u_{3i}\,q^{\frac{1}{2}}-\frac{q^{-\frac{1}{2}}}{u_{2j}\,u_{3i}})}
{(\frac{u_{3i}}{u_{2j}}q^{-\frac{1}{2}}-\frac{u_{2j}}{u_{3i}}q^{\frac{1}{2}})
(u_{2j}\,u_{3i}\,q^{-\frac{1}{2}}-\frac{q^{\frac{1}{2}}}{u_{2j}\,u_{3i}})}\ \ 
\prod_{j\neq i}^{M_{3}}
\frac{(\frac{u_{3i}}{u_{3j}}q^{-1}-\frac{u_{3j}}{u_{3i}}q)
(u_{3i}\,u_{3j}\,q^{-1}-\frac{q}{u_{3i}\,u_{3j}})}{(
\frac{u_{3j}}{u_{3i}}q^{-1}-\frac{u_{3i}}{u_{3j}}q)
(u_{3i}\,u_{3j}\,q-\frac{q^{-1}}{u_{3i}\,u_{3j}})}\ 
\nonu
i=1,\ldots,M_{3} \nonumber
\een
}
The transfer matrix eigenvalues are obtained from 
(\ref{eq:gammak}), remarking that $\Lambda(u^{(1)})=\wh \Gamma^{(1)}(u)$:
\ben
 \Lambda(u^{(1\dots \fm+\fn-1)})&=&\sum_{k=1}^{\fm+\fn}
\,\wt\fm_k(u^{(k+1\dots\fm+\fn-1)})\,\wh 
\Lambda_{k}(u^{(k+1\dots\fm+\fn-1)})\times \nonu 
&&\times \prod_{j=1}^{M_{k}} \ff_k(u^{(k+1\dots\fm+\fn-1)},u_{kj})\ 
\prod_{j=1}^{M_{k-1}} 
\wt \ff_k(u^{(k+1\dots\fm+\fn-1)},u_{k-1j}),\label{eq:Lambdafin}\\
\wt\fm_{\fm+\fn}(u)&=&(-1)^{[\fm+\fn]}m_{\fm+\fn}.\nonumber
\een
 The Bethe equations
(\ref{BE}) ensure that $\Lambda(u)$ is analytical, in accordance with
the analytical Bethe ansatz. 
The Bethe vectors take the form:
\ben
\Phi(\{u\}) &=&\wh B^{(1)}_{a_1}(u_{11}) \cdots 
\wh B^{(1)}_{a_{M_1}}(u_{1M_1})\,\wh F^{(1)}_{a_1\dots 
a_{M_1}}(\{u\})\,\Omega, 
\label{eq:Phi51}\\[1.2ex]
&=&\wh B^{(1)}_{a^1_1}(u_{11})\cdots 
\wh B^{(1)}_{a^1_{M_1}}(u^{(1)}_{M_1})
\,\wh {\wt B}^{(2)}_{a^2_2}(u_{21})\cdots\wh {\wt
B}^{(2)}_{a^2_{M_2}}(u_{1M_2})
\cdots\wh {\wt
B}^{(\fn+\fm-1)}_{a^{\fn+\fm-1}_{M}}(u_{\fn+\fm-1,M})
\,\Omega^{(\fn+\fm-1)}\,.\nonumber
\een
We recall the notation $M=\sum_{j=1}^{\fn+\fm-1}M_{j}$, 
$\Omega^{(k)}= \big(e^{(k-1)}_{1}\big)^{\otimes 
M_{k-1}}\otimes\Omega^{(k-1)}$, 
$\Omega^{(1)}=\Omega$ and the auxiliary 
spaces are indicated according to remark \ref{rmk:auxspace}. 

%%%%%%%%%%%%%%%%%%%%%%%%%%%%%%%%%%%%
\section{Bethe vectors\label{sec:betheV}}
We present here a generalization to open spin chains of the 
recursion and trace formulas for Bethe vectors, obtained in \cite{MTV,TV} (see also 
\cite{BR1}) for closed spin chains. To our knowledge, this 
presentation for open spin chain is entirely new.

\subsection{Recursion formula for Bethe vectors}

{From} expression (\ref{eq:Phi51}), we can extract a recurrent form 
for the Bethe vectors,
\ben
\Phi^{\fn+\fm}_M(\{u\}) &=& \wh B^{(1)}_{a^1_1}(u_{11}) \cdots 
 \wh  B^{(1)}_{a^1_{M_1}}(u_{1M_1})\,\wh\Psi^{(1)}_{\{u_1\}}
\Big(\Phi^{{\fn+\fm}-1}_{M-M_{1}}(\{u_{(>1)}\})\Big),
\label{BVR}\\
\wh\Psi^{(1)}_{\{u_1\}} &=& 
v^{(2)}\,\circ \,
(\tau \otimes \pi^{(2)}_{u_{1M_1}} \otimes \dots \otimes 
\pi^{(2)}_{u_{11}}) \circ \Delta^{(M_1)} ,\label{eq:recurPhi}
\een
where $\pi_{a}$ is the fundamental representation evaluation 
homomorphism normalized as in (\ref{eq:eval-k}), 
$v^{(k)}$ is the application of the highest weight vector 
$e_{1}^{(k-1)}$:
\begin{equation}
v^{(k)}(X\,\Omega^{(k-1)})=X\,(e_{1}^{(k-1)})^{\otimes 
M_{k-1}}\otimes\,\Omega^{(k-1)}=X\,\Omega^{(k)}\,,
\end{equation}
and $\tau$ is the morphism 
\ben
\tau: \begin{array}{lcl}  
\fD_{\fm-1|\fn} & \to & \fD_{\fm|\fn}/\cI_{1} \\[1.2ex]
d_{ij}(u) & \mapsto &   \wh d_{i+1,j+1}(u)\\
\end{array}.
\een
If we denote by $[.]_{\fm|\fn}$ the grading used in the 
$\fD_{\fm|\fn}$ superalgebra, the mapping $\tau$ corresponds to the 
identification $[j]_{\fm-1|\fn}=[j+1]_{\fm|\fn}$.

Remark that since all the operators in (\ref{BVR}) apply on the 
pseudo-vaccum, one can consider the operators built from $\tau$ as 
belonging to $\fD_{\fm|\fn}$ instead of 
$\fD_{\fm|\fn}/\cI_{1}$.
So by induction we build the Bethe vectors from $\fD_{\fm|\fn}$ 
generators and $R$-matrices
in auxiliary spaces. 

\subsection{Supertrace formula for Bethe vectors\label{sect:supertrace}}

We can also write the Bethe vector into a supertrace formula and 
prove the equivalence with the 
recursion relation discussed above.

\begin{theorem}\label{theo:traceForm}
The Bethe vector (\ref{BVR}) admit a supertrace formulation. 
We note $A_1; \dots; A_{\fm+\fn-1} $ the ordered sequence of 
auxiliary spaces
$a_1^1, \dots, a_{M_1}^1$; $a_1^2, \dots, a_{M_2}^2$
$; \dots ;$ $a_1^{\fm+\fn-1}, \dots ,
a_{M_{\fm+\fn-1}}^{\fm+\fn-1}$.
\beq
\Phi^{\fn+\fm}_M(\{u\}) = (-1)^{G_1}\,
str_{A_1 \ldots A_{\fm+\fn-1} }
\left(\prod_{i=1}^{\fm+\fn-1}
\wh {\DD}^{(i)}_{A_i}(\{u_i\})\,
 E_{{\fn+\fm},{\fn+\fm}-1}^{\vec{\otimes}M_{\fn+\fm-1}} 
\otimes \dots \otimes  
E_{21}^{\vec{\otimes}{M_1}}\right) \,\Omega, \quad
\label{eq:Phi-str}
\eeq
where
\ben
\wh {\DD}^{(i)}_{A_i}(\{u_i\})&=&\prod_{j=1}^{M_i}
\wb \cR^{(i)}_{A_{<i},a^i_j}(\{u_{i-1}\},u_{ij})\,
\wh{D}^{(i)}_{a^i_j}(u_{ij})\,
\cR^{(i)}_{a^i_j,A_{<i}}(\{u_{i-1}\},u_{ij})\,,
\\
G_k &=& \sum_{i=k}^{{\fn+\fm}-2}\frac{M_i(M_i+1)}{2}[i] \,,
\label{eq:defAk}
\een
\ben
\wb \cR^{(i)}_{A_{<i},a^i_j}(\{u_{i-1}\},u_{ij})
 &=&\prod_{b<i}^{\longrightarrow} 
\prod_{c=1}^{\atopn{\longrightarrow}{M_b}}
\wb \RR^{(i,b)}_{a_j^ia_c^b }(u_{ij},u_{bc}^{(b+1\dots i-1)})\,,
 \\
 \cR^{(i)}_{a^i_j,A_{<i}}(\{u_{i-1}\},u_{ij})
 &=&\prod_{b<i}^{\longleftarrow}
\prod_{c=1}^{\atopn{\longleftarrow}{M_b}}
 \RR^{(b,i)}_{a_c^ba_j^i } (u^{(b+1\dots i-1)}_{bc},u_{ij})\,.
\label{eq:bigR} 
\een 
\end{theorem}
\prf
Equivalence is proven along the following lines.
Starting from expression (\ref{eq:Phi-str}), we can extract the 
$M_1$ auxiliary spaces
corresponding to the first step of the nested Bethe ansatz :
\begin{eqnarray}
\Phi^{\fn+\fm}_M(\{u\}) &=& (-1)^{\frac{M_1(M_1+1)}{2}\,[1]}\, str_{A_1} 
\Big[\,\wh {\DD}^{(1)}_{A_1}(\{u_1\}) \times (-1)^{G_2}\times \nonu
&&str_{A_2 \ldots A_{\fm+\fn-1}}
\,\Big(\prod_{i=2}^{\fm+\fn-1}
\wh {\DD}^{(i)}_{A_i}(\{u_i\})\,  
E_{{\fn+\fm},{\fn+\fm}-1}^{\vec\otimes{M_{\fn+\fm-1}}} 
\otimes \dots \otimes E_{32}^{\vec\otimes{M_2}} \Big)\,
\otimes \, E_{21}^{\vec\otimes{M_1}} \Big] \,\otimes \Omega. 
\qquad\qquad\qquad \nonumber
\end{eqnarray}
Using the isomorphism
$End(\CC^{\fm+\fn}) \sim \CC^{\fm+\fn}\otimes \CC^{\fm+\fn}$,
one can rewrite, for any $A(v)$, the supertrace 
with an $E_{21}$ matrix as:
\begin{eqnarray}
str\Big(\wh D^{(1)}(u)\,A(v)\,E_{21}\Big) 
&=& \sum_{j=1}^{\fm+\fn} \big( e_{1}^t\otimes e_{j}^t\otimes \wh d^{(1)}_{1j}(u)\big)
\,A(v)\, \big(e_{1}\otimes e_{2}\otimes 1\big), 
\label{eq:str-form}
\nonu
&=&(-1)^{[1]+[1]\,[A]}\wh B^{(1)}(u)A(v)\, \big(e_{2}\otimes 1\big).\,
\end{eqnarray}
Using formula (\ref{eq:str-form}) for the auxiliary spaces 
$1,\ldots,M_{1}$, and remarking that the case 
$j_a=1$ for $a =1, \dots, M_1$ does not contribute, we obtain:
\ben
\Phi^{\fn+\fm}_M(\{u\}) &=& 
\wh B^{(1)}_{a^1_1}(u_{11})\cdots 
\wh B^{(1)}_{a^1_{M_1}}(u_{1M_1})\,(-1)^{G_{2}}\, \times \nonu
&&str_{A_2 \ldots A_{\fm+\fn-1}}
\,\Big(\prod_{i=2}^{\fm+\fn-1}
\wh {\DD}^{(i)}_{A_i}(\{u_i\})\,  
E_{{\fn+\fm},{\fn+\fm}-1}^{\vec\otimes{M_{\fn+\fm-1}}} 
\otimes \dots \otimes E_{32}^{\vec\otimes{M_2}} \Big) \,
\Omega^{(2)}.
\een
To end the proof, we remark that:
\begin{eqnarray}
(-1)^{G_{2}}\,str_{A_2 \ldots A_{\fm+\fn-1}}
\,\Big(\prod_{i=2}^{\fm+\fn-1}
\wh {\DD}^{(i)}_{A_i}(\{u_i\})\,  
E_{{\fn+\fm},{\fn+\fm}-1}^{\vec\otimes{M_{\fn+\fm-1}}} 
\otimes \dots \otimes E_{32}^{\vec\otimes{M_2}} \Big)\Omega^{(2)}=\,\wh\Psi^{(1)}_{\{u_1\}}
\Big(\Phi^{{\fn+\fm}-1}_{M-M_{1}}(\{u_{(>1)}\})\Big)\nonumber
\end{eqnarray}
which allows to recover the form (\ref{eq:recurPhi}).
\finprfbis

\begin{rmk}[Conjecture] Although theorem \ref{theo:traceForm} has been 
proven only when 
$K^{+}(u)$ belongs to a NABA couple, the expression (\ref{eq:Phi-str})
does not depend 
on $K^{+}(u)$: we conjecture that this expression 
 is valid for any couple of diagonal $K^{\pm}(u)$ matrices. 
This conjecture is supported by the fact that 
 analytical Bethe ansatz is known to work for 
any diagonal boundary matrices. 
\end{rmk}

\subsection{Examples of Bethe vectors}
To illustrate the supertrace formula, we present here some explicit 
examples of Bethe vectors asssociated to small numbers of excitations.

\paragraph{Bethe vectors of $\cD_{\fm|\fn}$ with $\fn+\fm = 2$ and 
$M_1=M$.}
We reproduce here the well-known case obtained with algebraic Bethe 
ansatz (see also section \ref{sec:ABA}). 
\ben
\Phi^{2}_M(\{u\}) = (-1)^{M [2]}\,\wh{d}^{(1)}_{12}(u_{11}) \cdots 
\wh{d}^{(1)}_{12}(u_{1M})\, \Omega.
\label{eq:Bethegl2}
\een
Note that this expression is also valid when $\fn+\fm>2$, setting 
$M_{1}=M$ and $M_{k}=0$, $k>1$.

\paragraph{Bethe vectors of $\cD_{\fm|\fn}$ with $\fn+\fm = 3$, $M_1=1$ 
and $M_2=1$.}

\begin{eqnarray}
\Phi^{3}_{1,1}(\{u\}) \ &=&\ (-1)^{[1]+[2]+[3]}\,
\frac{\fb(u_{11},u_{21})}{\fa_2(u_{11},u_{21})}  \,
\wh{d}^{(1)}_{12}(u_{11}) \,\wh{d}^{(2)}_{23}(u_{21})\,\Omega \nonu
&&+(-1)^{[1]+[2]+[3]}\,
\frac{\wb{\fb}(u_{21},u_{11})}{\wb{\fa}_2(u_{21},u_{11})}\, 
\frac{\fw_{32}(u_{11},u_{21})}{\fa_2(u_{11},u_{21})}  \,
\wh{d}^{(1)}_{13}(u_{11}) \,\wh{d}^{(2)}_{22}(u_{21})\,\Omega \nonu
&&+(-1)^{[1]}\,
\frac{\wb{\fw}_{23}(u_{21},u_{11})}{\wb{\fa}_2(u_{21},u_{11})} \,
\frac{\fb(u_{11},u_{21})}{\fa_2(u_{11},u_{21})}\,\wh{d}^{(1)}_{13}(u_{11}) 
\,\wh{d}^{(2)}_{33}(u_{21})\, \Omega. 
\nonumber
\end{eqnarray}
Again, this expression is also valid when $\fn+\fm>3$, setting 
$M_{k}=0$, $k>2$.

\null

We also computed the Bethe vectors corresponding to  
$M_1=M_2=M_3=1$ and $M_{k}=0$, $k>3$. Their expression is rather 
long, with 11 different terms: 
 we do not write it here explicitly.

\section{Conclusion}

In this paper, we have proposed a global treatment of the NBA for 
\textbf{universal transfer matrices} 
of open spins chains with
\textbf{NABA couple} of boundary matrices. The modification of the 
nested Bethe ansatz appliable to diagonal boundary matrices that do 
not form a NABA couple remains to be found. Since the analytical Bethe 
ansatz can be performed in this case, such a refinement should be 
possible.

We have computed a \textbf{trace formula} for the Bethe vector of the 
open chain. This formulation could be a starting point for the investigation 
of the quantized Knizhnik-Zamolodchikov equation following the work \cite{TV2}.
For such a purpose, the coproduct properties of Bethe vectors for open 
spin chains remain to be studied.
Defining a scalar product and computing the norm of these Bethe 
vectors is also a point of  fundamental interest.

From a different point of view, this trace formula and the mapping between the reflection 
algebras of different size
could be the starting point for the construction of a Drinfeld's current 
realisation \cite{D2} for the reflection algebra in the spirit of \cite{KhoPak} 
on the current realisation of the Bethe vector for the periodic case.

The case of open spin chains with  general boundary matrices is also 
a subject of fundamental interest. A deeper understanding of representations of 
reflection algebras when the $K$ matrix is not diagonal may be of 
some help. Alternatively, a different approach using another 
presentation of the reflection algebra could be the clue to go beyond 
the results obtained so far. Some works have been done for the $\fm+\fn=2$ case 
in \cite{BK}, but  
the general treatment  for universal transfer matrices remains an 
open problem. 
The functional approach developped in \cite{TQ} for 
$\fm+\fn=2$ also deserves a generalization, both for universal 
transfer matrices, and for bigger algebras.  

\pagebreak[0]
\appendix
\section{R and M matrices \label{appendix1}}
We remind the general form of the $R$-matrices we used in the paper \cite{BR1}.
Note we use a more compact form:
\ben
R_{12}(u,v) &=& \fb(u,v)\,\II \otimes \II
+ \sum_{i,j=1}^{{\fm+\fn}} \fw_{ij}(u,v)  E_{ij}\otimes E_{ji}, \\
\fw_{ij}(u,v)&=&\begin{cases}
\fa_i(u,v)-\fb(u,v) \mb{for} i=j\\
\fc_{ij}(u,v) \mb{otherwise}
\end{cases}, \nonu
\bar{R}_{12}(u,v) &=& R_{12}(u,\inv(v))= \bar \fb(u,v)\II \otimes \II
+ \sum_{i,j=1}^{{\fm+\fn}} \bar \fw_{ij}(u,v)  E_{ij}\otimes E_{ji}.
\een
The functions involved in these expressions are given by (with the 
convention $Y(\fm)\equiv Y(\fm|0)$, 
$\wh\cU_{q}(\fm)\equiv\wh\cU_{q}(\fm|0)$ and $[b]=0$, $\forall\,b$, in these 
two cases):

\paragraph{{For} $Y(\fm|\fn):$}
\begin{eqnarray}
{\mathfrak b}(u,v)&=& u-v 
\mb{;} \fa_{a}(u,v)= u-v-(-1)^{[a]}\,\hbar
\mb{and}
{\mathfrak w_{ab}(u,v)} = -(-1)^{[b]}\,\hbar 
\label{eq:abc-yang1}\\
{\bar{\mathfrak b}}(u,v)&=& u+v 
\mb{;} \bar\fa_{a}(u,v)= u+v-(-1)^{[a]}\,\hbar
\mb{and}
{\bar{\mathfrak w}_{ab}(u,v)} = -(-1)^{[b]}\,\hbar 
\label{eq:abc-yang2}
\end{eqnarray}

\paragraph{{For} $\wh\cU_{q}(\fm|\fn):$}
\begin{eqnarray}
{\mathfrak b}(u,v)&=&\frac{u}{v}-\frac{v}{u} 
\mb{;} \fa_{a}(u,v)= \frac{u}{v}\,q^{1-2[a]}-\frac{v}{u}\,q^{2[a]-1}
\label{eq:abc-Uq1}\\
\mbox{and}&&
\displaystyle{\fw_{ab}}(u,v)
\ =\ (-1)^{[b]}\,(q-q^{-1})\,
\left(\frac{u}{v}\right)^{sign(b-a)}\,,\ a\neq b
\een
\ben
{\bar{\mathfrak b}}(u,v)&=&u\,v-\frac{1}{u\,v}
\mb{;} \bar\fa_{a}(u,v)= u\,v\,q^{1-2[a]}-\frac{1}{u\,v}\,q^{2[a]-1}
\\
\mbox{and}&&
\displaystyle{\bar \fw_{ab}}(u,v) \ =\  
(-1)^{[b]}\,(q-q^{-1})\,(u\,v)^{sign(b-a)}\,,\ a\neq b
\label{eq:abc-Uq3}
\end{eqnarray}

The matrix $M$ is a diagonal matrix:
\ben
M=\sum_{i=1}^{\fm+\fn} m_i E_{ii}\mb{with}\begin{cases}
m_i=1 & \mb{for}Y(\fm|\fn)\\ 
m_i= q^{\fm-\fn-2k+1}\ q^{-2[k]+4\sum_{i=1}^{k} [i]}
& \mb{for}\wh\cU_{q}(\fm|\fn)
\end{cases}.
\een  

\section{Functions appearing in NBA\label{app:fct}}
The functions are constructed from the three functions appearing in 
the $R$-matrix, whose explicit form are given in equations 
(\ref{eq:abc-yang1})-(\ref{eq:abc-Uq3}) above:
\ben
\begin{array}{llcllcll}
\displaystyle\ff_i(u,v)=
\frac{\fa_i(v,u)\,\bar{\fb}(u^{(i)},v^{(i)})}{\fb(v,u)\,\bar{\fb}(u,v)} 
&,\qquad \displaystyle\wt{\ff}_{i+1}(u,v)=
\frac{\fa_{i+1}(u,v)\,\bar{\fa}_{i+1}(u,v)}{\fb(u,v)\,\bar{\fb}(u,v)} \\
\\
\displaystyle\fg_i(u,v)= 
\frac{\fc_{i-1i}(u,v)\,\bar{\fb}(v^{(i)},v^{(i)})}{\fb(u,v)\,\bar 
\fb(v,v)}
&,\qquad \displaystyle\wt \fg_{i+1}(u,v)=
-(-1)^{[i]+[i+1]} \frac{\fc_{k+1k}(u,v)\,\bar{\fa}_{k+1}(u,u)}{\fb(u,v)\,\bar\fb(u,u)} \\
\\
\displaystyle\fh_i(u,v)= 
-\frac{\bar{\fc}_{i-1i}(u^{(i)},v^{(i)})}{\bar{\fb}(u,v)}
&,\qquad \displaystyle\wt \fh_{i+1}(u,v)=
\frac{\bar \fc_{i+1i}(u^{(i)},v^{(i)})\,\bar \fa_{i+1}(u,u)\,
\bar \fb(v^{(i)},v^{(i)})}{\bar \fb(u,u) \,
\bar \fb(v,v)\,\bar \fb(u,v)}
\end{array}
\een
We also use (presented here for $\cA_{\fm|\fn}=\wh\cU_{q}(\fm|\fn)$; 
for $\cA_{\fm|\fn}=Y(\fm|\fn)$ one has to set $q=1$ in the relations 
below):
\ben
\psi_{i}(u) &=& \frac{\bar \fc_{i+1i}(u,u)}{\bar\fa_i(u,u)}
\label{eq:psij}\\
\chi_k(u) &=&\begin{cases}
& -q^{2[k]-1}\frac{\bar \fb(u,u)}{\bar \fb(u^{(k)},u^{(k)})} \eta_k(u,c_+) \mb{for} k=1 
\mb{and} a=1\\
&-q^{2[k]-1}\frac{\bar \fb(u,u)}{\bar \fb(u^{(k)},u^{(k)})} \mb{else}
\end{cases}\\
\eta_k(u,c_+) &=&\frac{\fb(\wt c_+,u^{(k+1)})}{\wb \fb(\wt c_+,u^{\wb{(k+1)}})}\\
\fe_k(u)\II^{(k)}_b &=&
tr_a(M_a^{(k)} \bar{\RR}^{(k)}_{ab}(u,u)\RR^{(k)}_{ba}(u,u))\nonu
&=&q^{-k+1}q^{-2\fn-2\sum_{i=k}^{\fm+\fn}[i]+4\sum_{i=1}^{\fm+\fn}[i]}(-1)^{[k]}
\frac{\bar\fb(u^{\wb{(k\dots \fm+\fn)}},u^{\wb{(k\dots \fm+\fn)}})}
{\bar\fa_{k}(u,u)}\II^{(k)}_b
\label{sol}
\\
\wt \fm_k(u)&=&q^{1-2[k]}\frac{\bar{\fa}_{k+1}(u,u)\,\fe_{k+1}(u)}
{\bar{\fb}(u,u)} \mb{for} k\neq1.
\een

The following useful relations are used in the paper:
\ben
\fb(u,v)\,
\fa_i(u^{\wb{(j)}},v^{\wb{(j)}})
&=&\fa_i(u,v)\,\fa_j(u,v)-\fw_{ij}(u,v)\,\fw_{ji}(u,v)
\,,\quad i \neq j,
\label{eq:fct-id1}\\
\fb(u,v)\,\fw_{ij}(u^{\wb{(k)}},v^{\wb{(k)}})&=&
\fw_{ij}(u,v)\,\fa_k(u,v)-\fw_{ik}(u,v)\,\fw_{kj}(u,v)
\,,\quad i > j >k;
\nonu
\fb(u^{\wb{(i)}},v^{\wb{(i)}})&=&\fa_i(u,v),
\\
\fa_j(u,v)-\fw_{ij}(u,v)&=&
\fb(u,v)\,q^{siq(j-i)(-1+2[j])} ,
\label{eq:fct-id2}
\\
q^{2-4[i]}\,\psi_{i-1}(u^{(i-1)}) &=& \psi_{i-1}(u^{(i-1,i)}) - 
\psi_{i-1}(u^{(i-1,i)})\,\psi_{i}(u^{(i)})
\label{eq:id-psi}
\een

% \section{biblio}

\end{document}